%% file: conference-cgo.tex
\documentclass[conference]{IEEEtran}
\IEEEoverridecommandlockouts

\usepackage{amsmath,amssymb,amsfonts}
\usepackage{graphicx}
\usepackage{textcomp}
\usepackage{xcolor}
\usepackage{comment}
\usepackage{subcaption}
\usepackage{wrapfig}
\usepackage{array}
\usepackage{booktabs}
\usepackage{makecell}
\usepackage{listings}
\usepackage[numbers]{natbib}
\usepackage{tikz}
\usepackage{svg}
\usepackage{balance}

\usepackage[normalem]{ulem}

\usetikzlibrary{matrix,positioning,arrows.meta} 

\input{prelude.tex}

\definecolor{LightGray}{gray}{0.9}

\def\BibTeX{{\rm B\kern-.05em{\sc i\kern-.025em b}\kern-.08em
    T\kern-.1667em\lower.7ex\hbox{E}\kern-.125emX}}    
\begin{document}

\newcommand{\revised}[1]{#1}
\newcommand{\amirbijective}[1]{#1}

\title{LEGO: A Layout Expression Language for Code Generation of Hierarchical Mapping}

\author{\IEEEauthorblockN{Amir Mohammad Tavakkoli}
\IEEEauthorblockA{Kahlert School of Computing \\
University of Utah\\
Salt Lake City, USA \\
amir.tavakkoli@utah.edu}
\and
\IEEEauthorblockN{Cosmin E. Oancea}
\IEEEauthorblockA{Dept. of Computer Science \\
University of Copenhagen\\
Copenhagen, Denmark \\
cosmin.oancea@di.ku.dk}
\and
\IEEEauthorblockN{
Mary Hall}
\IEEEauthorblockA{Kahlert School of Computing\\
University of Utah\\
Salt Lake City, USA \\
mhall@cs.utah.edu}
}

\maketitle

\begin{abstract}
We describe LEGO, a new approach to optimizing data movement whereby code is expressed as a layout-independent computation and composed with layouts for data and computation. 
This code generator organization derives complex indexing expressions associated with hierarchical parallel code and data movement for GPUs.  LEGO maps from layout specification to indexing expressions, and can be integrated into existing compilers and code templates.  It facilitates the exploration of data layouts in combination with other optimizations.  We demonstrate LEGO's integration with the Triton and MLIR compilers, and with CUDA templates.  We show that LEGO is capable of deriving 
performance competitive with Triton, and shows broad applicability \revised{for data and thread layout mapping optimizations} in its integration with~CUDA~and~MLIR.  
\end{abstract}


\section{Introduction}

Now that Moore's Law and Dennard scaling no longer drive performance improvements, researchers have turned to architecture specialization and domain-specific programming systems for further scaling gains~\cite{Leiserson20}.  Data movement is now the dominant cost in execution time and energy~\cite{Kogge13}, and optimizations to reduce data movement must take center stage.  


Most commonly, optimizing data movement involves \textit{reordering computation} to modify memory access order; this reordering allows the computation to exploit reuse of data in nearby \textit{fast} memory, especially cache and registers using loop transformations such as tiling~\cite{WolfeTiling} and unroll-and-jam~\cite{Carr}.  \revised{Notably, polyhedral compiler frameworks, dating back to mid 1980s~\cite{Feautrier92} -- and more recently Pluto~\cite{pluto}, PPCG~\cite{PPCG}, and Polygeist~\cite{polygeist}, among others -- represent a dynamic instance of a statement in a multi-dimensional loop iteration space as an integer point in the statement's polyhedron~\cite{polyhedralsurvey}.  This mathematical representation facilitates the composition of complex transformation sequences as statement instance reorderings.}

As an alternative to statement reordering, a system can \textit{change the layout of data in memory} to more closely match the order in which the computation accesses it.  For example, the standard layout for a 2-dimensional array in a C or C++ compiler is \text{row-major order}, whereby adjacent elements in a row are stored contiguously in memory, and elements in the same column are strided by the length of the row.  But improved spatial reuse and reduced data movement has been demonstrated by alternatives to row-major order
~\cite{wise2001language,chatterjee1999nonlinear,bricks,Araya-Polo:2009:SIT:1507443.1507449,yount2016yask}. Further, controlling data layout helps achieve performance portability across architectures, matching layout to size and bandwidth of each architecture's memory hierarchy~\cite{kokkosview2024,Unat2016,DISTAL,Legion,brick-journal}.  

Recently, a body of work targeting vector and matrix processors in GPUs uses \textit{both data and computation layout} as well as data movement specification to decompose computations and order data to 
match the inputs and outputs of these accelerator units, e.g.,  
Fireiron~\cite{fireiron}, CuTe~\cite{CuTE}, Graphene~\cite{hagedorn2023graphene} and Triton~\cite{triton,triton2}. 
%
%
Such systems {\em restrict} indexing expressions to 
encode \textit{linear} formulas that are represented in terms of \textit{strides} or \textit{binary matrices}, which makes specification tedious and error-prone.  
\amirbijective{Despite the availability 
of implementations of some of these, 
integration into other tools and generalization 
for other architectural features remain limited.}

This paper presents LEGO, a layout abstraction that increases generality and facilitates adoption of this essential capability by other frameworks.
LEGO can express any bijective mapping between the logical and reordered index space that are represented as {\em permutations}, thus omitting strides. Permutations can be {\em linear}, e.g., of (entire) dimensions, or {\em irregular}, represented by user-defined functions. 
LEGO’s lack of explicit strides eliminates low-level index calculations, making code simpler and more expressive than frameworks like Triton. Its high-level building blocks reduce both mathematical complexity and overall code size, allowing users to modify computations simply by changing layouts without altering core logic. 
Once the layouts and connections are defined, LEGO automatically generates a \amirbijective{bijective} mapping for the whole ensemble, thus serving both as a high-level programming abstraction and a tool for high-performance code generation. \amirbijective{
LEGO's algebra is also extended 
beyond this bijective restriction to support partial tiles and common injective mappings.}

This paper makes the following contributions:

\begin{figure*}[!ht]
\centering
\includegraphics[width=0.9\textwidth]{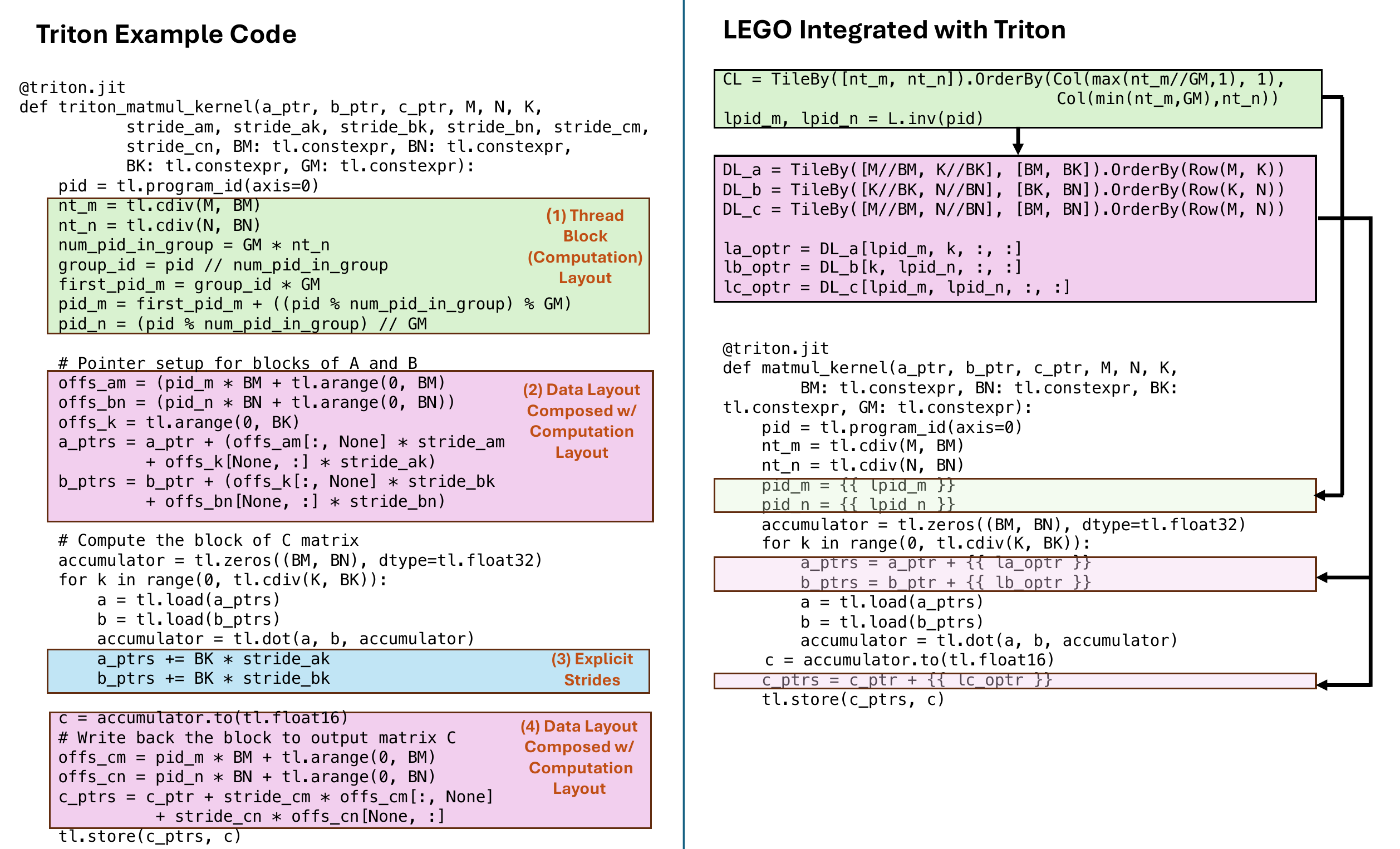}
\caption{Matrix multiplication expressed in Triton (left) and using LEGO to instantiate Triton (right).  The latter version describes layouts for thread block and data at a high level; LEGO automatically derives the complex indexing expressions.  The resulting code generated by LEGO is shown in Figure~\ref{fig:triton_output}.} 
\vspace*{-.2in}
\label{fig:lego_triton_motivation}
\end{figure*}

\begin{itemize}
\item a general 
abstraction for \amirbijective{bijective} layouts, which can express both computation and data, 
\item an implementation 
reproducible from the paper and accessible to other frameworks via open-source software,  
\item a demonstration of lowering to Triton, CUDA,~and~MLIR supporting irregular layouts such as an anti-diagonal,
\item a demonstration of ease of use and performance competitive with state of the art. 
\end{itemize}

\revised{Layout algebras such as LEGO are an important component of emerging compiler frameworks with data/thread tile abstractions, and can facilitate layout exploration.  To fully exploit data/thread tiles requires integration with statement reordering optimizations such as the previously-described polyhedral compilers -- a substantial implementation effort. In this paper, we demonstrate
LEGO's use as a frontend tool to generate optimized indexing expressions automatically from data and thread layout specifications, and rely on backend compiler frameworks to perform loop iteration space reorderings.    
}




\section{Motivation and Overview}
\label{sec:motivation}
In this section, we discuss how layout descriptions are being used in existing tools to simplify the development of high-performance GPU implementations of matrix multiply, using a tiled, hierarchical approach to achieve data locality in registers and shared memory, and leveraging matrix processors.  We
motivate LEGO's generalization of the specification of the layout and automatic generation of the indexing expressions.  

\paragraph{Matrix Multiplication Using Triton}  

High-performance implementations of matrix multiply for GPUs can be achieved with Triton~\cite{triton} programs, as shown in Figure~\ref{fig:lego_triton_motivation}~(left).
The Triton program calculates the memory offsets for the input and output matrices, loads the necessary elements from \( A \) and \( B \), and subsequently performs the dot product to compute the matrix multiplication, storing the result in \( C \). 
In the example, the code describes 2-D tiles to reuse inputs and produce output tiles.  
The compiler detects these tiles and optimizes the \texttt{load} and \texttt{store} operators to move data through the GPU memory hierarchy and generates the code for the \texttt{dot} operator. 
As a result of the compiler's careful management of the memory hierarchy and mapping to Tensor Cores, this implementation demonstrates competitive performance with the cuBLAS library. 


To derive the Triton program in Figure~\ref{fig:lego_triton_motivation}~(left) nevertheless requires the programmer to write complex code, enclosed in colored boxes, to express computation and data layout specifications, specifically: (1) the thread-block-level computation layout (green); 
(2) the layout in global memory of the input matrices \( A \), and \( B \) and hints at their 2D tiles, composed with the computation layout (pink); (3) an explicit stride for \( A \), and \( B \) (blue); and, (4) the layout in global memory of output matrix \( C \) (pink).  In particular, the thread block layout in the green box is non-standard and has been found to perform better than a 2D row-major order.  At the inner level, program IDs are grouped with a group size of $GM$, while the outer level defines the overall ordering of these groups, with both levels using a column-major order.
Not only is a significant portion of the code dedicated to complex index calculations, but  
the implementation is
tightly coupled to the program instance layouts, fixed $K$ iteration space layout, and the data layout of matrices \( A \), \( B \), and \( C \). 

\paragraph{Matrix Multiplication Using Graphene} 

Graphene~\cite{hagedorn2023graphene} is an intermediate representation for specifying data layout and data movement using the shape algebra of CuTe~\cite{CuTE}, a part of NVIDIA's CUTLASS library.  Graphene improves upon the interface for Triton by: (1) supporting more general data layouts of strided rectangular regions, as specified using a shape algebra; and (2) generating the complex index expressions automatically through a mapping from the shape algebra.  A performance engineer writes a template in the Graphene IR that expresses data and thread layouts, which is instantiated by the Graphene compiler. A simple Graphene template for Matrix Multiplication takes 
22 lines of specification (see Figure~8 in reference~\cite{hagedorn2023graphene}).



\paragraph{LEGO Improvements}
Like Graphene, LEGO derives index expressions from layout specifications, freeing the programmer from providing these low-level details. 
As compared to Graphene, LEGO eliminates explicit stride specifications in the layout definition (section~\ref{subsec:comparison-graphene}) and extends support to any
bijective mapping from multidimensional coordinates to contiguous linear space, an aspect not supported by previous work. 
In Figure~\ref{fig:lego_triton_motivation}~(right), at the top, we show LEGO specification for thread block layout (2D column major order shown in green box). Next, the data tiles and their composition with the thread-block layout in row major for the input and output matrices are specified (pink box). The Triton kernel is now much simpler; LEGO instantiates the thread-block code and computes data addresses, reducing the number of arithmetic operations the user must specify from 31 to just 9.

Moreover, LEGO is a building block for compilers or code generators that applies to computations beyond tensors or specific tensor cores. 
This point is demonstrated with CUDA code, the Triton and MLIR compilers (sections~\ref{sec:Implementation} and~\ref{sec:results}). 

\section{LEGO Specification}
\label{sec:lego-spec}

%

The discussion is organized as follows: section~\ref{subsec:math-intuition}
presents in an intuitive fashion how the LEGO pieces are composed, 
section~\ref{subsec:lego-blocks} shows the LEGO grammar and uses it
to define the semantics of a LEGO ensemble from the semantics of 
individual pieces, and section~\ref{subsec:comparison-graphene} 
compares the expression of CuTe/Graphene and LEGO layouts.

\subsection{LEGO by Example \& Mathematical Intuition}
\label{subsec:math-intuition}

LEGO elevates data layout to a first-class design
consideration. 
The user defines a logical view of the
index space together with reordering transformations, which can
be (de)composed hierarchically and chained horizontally.

\begin{figure}[t]
\includegraphics[width=.99\linewidth]{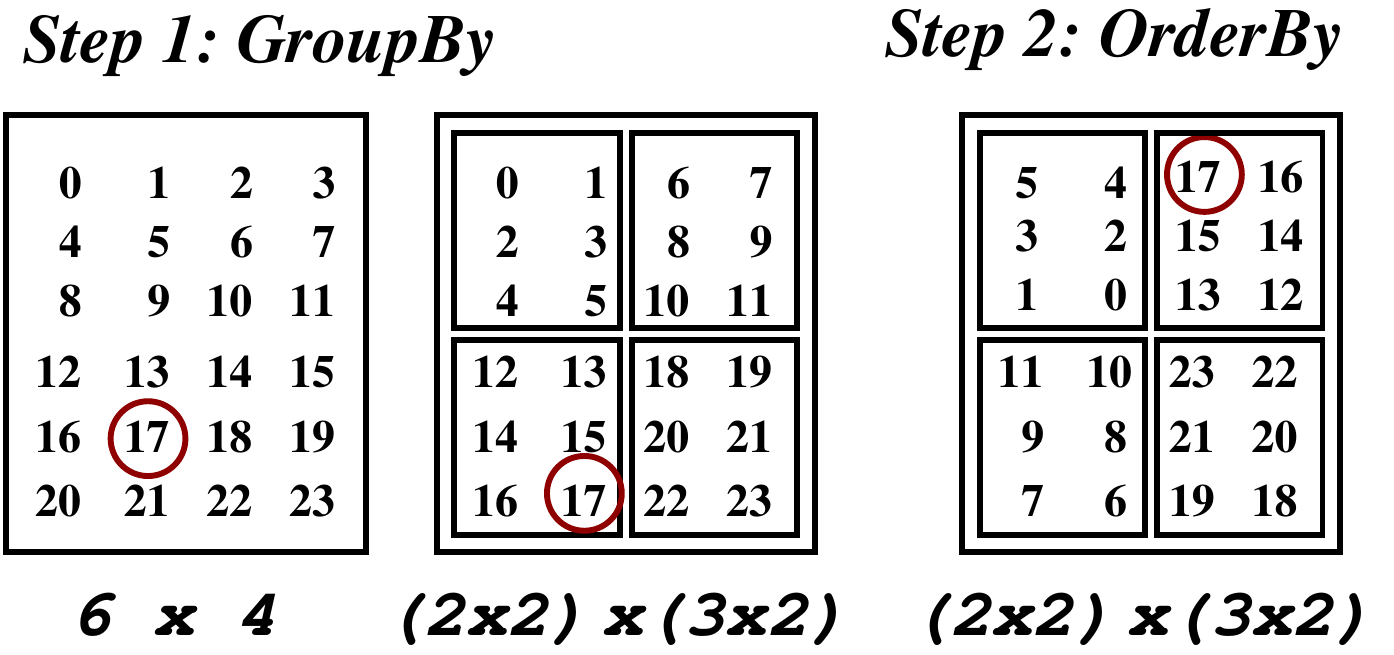}
\vspace{-1ex} 
\caption{
Logical view~--~reshape~--~permute, hierarchically. 
}
\label{fig:lego-eg1}
\end{figure}

Figure~\ref{fig:lego-eg1} demonstrates a simple use case that
defines a LEGO layout for a flat buffer consisting of $N=24$
elements.   The user specifies the logical view of
a multi-dimensional array 
as part of expressing
the target algorithm.   This is shown in the left column as
an array $A$ of shape $6\times 4$, whose elements correspond
to the flat index space of the logical view, e.g.,
$A[4,1] = 4\cdot 4 + 1 = 17$. 

Next, the user would like to reorder in memory the elements
of the logical view. The first step in this process is to
define a hierarchy of some $q$ levels of tiles, each of the
same dimensionality $d$:
\begin{equation}\label{eq-hier-tiles}
N = (n^1_1\times\ldots\times n^1_d)\times\ldots\times(n^q_1\times\ldots\times n^q_d)
\end{equation}
\noindent 
The middle column of Figure~\ref{fig:lego-eg1} demonstrates
this step for $q=2$ and $d=2$, creating a 4D array whose
outer and inner tiles have shapes $2\times 2$ and 
$3\times 2$, respectively:
\[
N = (n^1_1 \times n^1_2) \times (n^2_1\times n^2_2) = (2\times 2) \times (3\times 2)
\]
\noindent{}Note that this step is just a reshape operation 
applied to the logical layout that does not change yet the physical
layout, i.e., laying down the elements of the array in the middle
column in increasing order of inner dimensions still results in 
$[0, 1, \ldots , 22, 23]$.
 
The second step reorders the elements of tiles
by defining permutations (possibly) at each of the $q$ levels
of~the~hierarchy:
{\em The general case} is covered by a pair of user-defined functions 
implementing a bijection between the index space of the corresponding
tile and its  
flat space. For example, the right column
of the figure shows that elements of each innermost tile are
reordered according to the user-defined permutation
$p_{n^2_1,n^2_2}(i,j) = (n^2_1\text{-}1\text{-}i)\cdot n^2_2 + (n^2_2 \text{-} 1 \text{-} j)$,
which reverses the elements on each of the two dimensions.

For ease of use, LEGO also supports a 
{\em specialized case} that interchanges the dimensions of a tile
by some statically known permutation $\sigma$ of $[1,\ldots,d]$.
%
%
The right column of Figure~\ref{fig:lego-eg1} uses $\sigma = [2,1]$
on the outermost tile level to transpose the locations of the
inner tiles. 
Such a reordering allows the user to bypass the 
hassle of writing functions, and may enable further 
simplifications of index computation and analyses. 
The layout transformation of 
Figure~\ref{fig:lego-eg1} is expressed as:

\vspace{-2ex}
\begin{footnotesize}
\[
\textbf{GroupBy}([6,4],\textbf{OrderBy}(\textbf{RegP}([2,2], [2,1]),\textbf{GenP}([3,2], p, p^{-1}))))
\]
\end{footnotesize}
\vspace{-2ex}

\noindent
where $[6,4]$ specifies the shape of the logical view. The \textbf{OrderBy} construct specifies 
that the outer tile level,
of shape $[2,2]$, is reordered by transposing its dimensions, hence
$\sigma = [2,1]$, and the elements of the inner tiles, of
shape $[3,2]$, are reordered by the user-defined permutation $p$
(whose inverse $p^{-1}$ is not shown).

The user interface with the layout consists of two functions: 
(1) \texttt{apply}, which maps a logical-view index to its flat physical
position; and, (2) \texttt{inv} that performs the reverse. For example, 
$\texttt{apply}([4,1]) = 6$ and $\texttt{inv}(6) = [4,1]$. 
The rationale is that element $17$ at index $[4,1]$ in the logical
view is ultimately placed in memory at position~$6$, corresponding
to index $[0,1,0,0]$ of the 4D array shown~on~the~right:
%
\begin{itemize}
\item $17$ belonged to tile $[1,0]$ of the array in the middle column
      and transposition has brought its tile to position $[0,1]$
\item within its tile, $17$ was placed as the last index of
      both dimensions, and reverting them brings
      it in position $[0,0]$. 
\end{itemize}


\paragraph{Mathematical Intuition}
The mathematical glue that binds the multi-layered components
are the well known {\em canonical bijections}, denoted 
$\mathcal{B}$ and $\mathcal{B}^{-1}$, that connect a
multi-dimensional index space to its corresponding flat space.
For example, the index transformation between the logical view
and the reshaped tile hierarchy---i.e., between the left and
middle columns of Figure~\ref{fig:lego-eg1}---is obtained by
(1) flattening the logical-view index by applying $\mathcal{B}$,
and then by (2) unflattening the resulted index in the tiled
space by applying $\mathcal{B}^{-1}$.

LEGO enables the user to express piece-wise bijections that
document the reordering performed at each level of the tile
hierarchy, and provides an automatic procedure that combines
these into one bijection \textbf{B} that covers the whole
index space. 
This essentially allows the user to work in a suitable logical
space, say of shape $n'_1\times\ldots\times n'_{d'}$, while
LEGO transparently performs the mapping to the reordered flat
(physical) space by means of the \texttt{apply} bijection:
\[
\mathbf{B} \circ \mathcal{B}^{-1}_{(n^1_1\cdot\ldots\cdot n^1_d)~\cdot~\ldots~\cdot~(n^q_1\cdot\ldots\cdot n^q_d)} \circ \mathcal{B}_{n'_1,\ldots,n'_{d'}}
\]
with $n^1_1\ldots n^q_d$ defined in Equation~\cref{eq-hier-tiles}.
As well, since bijections are reversible, one can find the 
logical multi-dimensional index corresponding to a physical
one by using the \texttt{inv} bijection, inferred as:
\[
\mathcal{B}^{-1}_{n'_1,\ldots,n'_{d'}} \circ \mathcal{B}_{(n^1_1\cdot\ldots\cdot n^1_d)~\cdot~\ldots~\cdot~(n^q_1\cdot\ldots\cdot n^q_d)} \circ \mathbf{B}^{-1}
\]
Finally, LEGO allows to chain reordering \textbf{OrderBy} transformations, 
by similarly gluing them with canonical bijections. The next section~\ref{subsec:lego-blocks} presents the LEGO grammar and
details its implementation.

\subsection{Building Blocks \& Lowering Algorithm}
\label{subsec:lego-blocks}

\begin{figure}
\begin{scriptsize}
  \hspace{-5ex} 
  \begin{minipage}[t]{0.72\linewidth}
    \[ 
    \begin{array}{l}
    \begin{array}{cclr}
    \sigma_d & ::= & [\overline{k}^d] & \hsp\hsp\hsp\hsp 
                                        \text{Ct.~Perm.~of~$[1\ldots d]$}\smallskip\\
    Tile_d  & ::= & [\overline{e}^d] & \text{Sizes of a $d$-dim tile}
    \end{array}\medskip\\
    \begin{array}{ccl}
    \hspace{-1ex}Prm_d & ::= & \textbf{RegP}(Tile_d, \sigma_d) \hsp\hfill
    \text{Regular Perm}\medskip\\ 
        & ~\mid~ & \textbf{GenP}(Tile_d, f, f^{inv}) \hsp\hfill\text{Irregular Perm} 
    \end{array}\medskip\\
    \begin{array}{ccl}
    OrderBy & ::= & \hspace{-2ex}\textbf{OrderBy}(\overline{Prm_d}^{q_2})
    \medskip\\
    GroupBy & ::= & \hspace{-2ex}\textbf{GroupBy}(\overline{Tile_{d'}}^{q_1}, ~\overline{OrderBy}^{q_3})
    \end{array} 
    \end{array}
    \]
  \end{minipage}\begin{minipage}[t]{0.27\linewidth} 
    \[
    \begin{array}{l}
    \begin{array}{ccl}
    e &  ::= & k \hsp\hfill \text{Ct.$\in \mathbb{Z}$}\\
      &~\mid~& x \hsp\hfill \text{Var.}\\
      &~\mid~& e + e  \hsp\hfill \text{Add}\\
      &~\mid~& e * e  \hsp\hfill \text{Mul.}\\
      &~\mid~& \ldots \hsp\hfill \text{Other}
    \end{array}\vspace{1ex}\\\hspace{2ex}
    \textbf{Notation:}\\\hspace{2ex}
    \begin{array}{l}
    q,d \hsp\hfill \text{sequence size}\\
    h,k   \hsp\hfill \text{seq. iterators}\\
    i,j   \hsp\hfill \text{indices}\\
    n,m   \hsp\hfill \text{int expressions} 
    \end{array}
    \end{array}
    \]
  \end{minipage}
\end{scriptsize}
  \medskip\\
\begin{small}
\textbf{Notation}: $\overline{o}^q$ is a sequence $o_1,\ldots,o_q$ of $q$ objects of
some kind.\\\hspace{13ex} 
$d'$ and $d$ denote the dimensionality of a tiling hierarchy.
\end{small}
\vspace{-2ex}\\
\caption{ Grammar: $GroupBy$ gives the logical view of
          an~index space whose elements are reordered
          by a chain of $OrderBy$.}
\label{fig:lego-grammar}
\end{figure}


Figure~\ref{fig:lego-grammar} presents the LEGO grammar:
A \textbf{GroupBy} consists of 
(1) a hierarchical tile decomposition on some arbitrary but
    fixed number $q_1$ of levels, such that each tile has
    the same dimensionality $d'$, together with
(2) a chain of reordering \textbf{OrderBy} transformations.

An \textbf{OrderBy} defines its own $d$-dimensional tile hierarchy
on some $q_2$ levels by means of a sequence of permutations $Prm_d$. 
$Prm_d$ has two constructors: 
$\textbf{GenP}$ denotes a general permutation of the {\em elements}
of a tile, by a user-defined function $f$, whose inverse is $f^{inv}$. 
$\textbf{RegP}$ denotes a regular (constant) permutation
$\sigma$ of the~{\em dimensions}~of~a~tile, i.e., if the logical shape
of the tile is $\overline{n}^d$ then the reordered shape is 
$\sigma(\overline{n}^d)$. Finally, a tile is represented by its shape,
as a list of dimension sizes, which are expressions. 

Of course, the total number of elements of the hierarchical tiling
defined by \textbf{GroupBy} must equal that of each of the chained \textbf{OrderBy}s.
In practice, tiles within an \textbf{OrderBy} or \textbf{GroupBy} do not have to
share the same dimensionality, e.g., one may use a $1$-D grid of
$3$-D blocks; we use this restriction to simplify the~presentation.

\begin{figure}
\begin{small}
\textbf{Notation}: $o_k$ denotes $k^{th}$ object from sequence $\overline{o}^q = o_1,\ldots,o_q$ and
        \\\hsp\hsp\hsp
        $\overline{o}^{h=q_1\ldots q_2}$ creates a new sequence from objects $o_{q_1},\ldots,o_{q_2}$.
  \hspace{-7ex}
    \[ 
    \begin{array}{l}
    \sigma^{-1}_d \ \text{is obtained by scattering $[1,\ldots,d]$ at the positions of $\sigma_d$.}
    \medskip\\
    \mathcal{B}_{\overline{n}^q}(\overline{i \ }^q) \ = \ i_1\cdot \prod_{k=2}^{q} n_k \ 
      + \ \ldots \ + \ i_{q-1}\cdot n_{q} \ + \ i_{q}\medskip\\
    \mathcal{B}^{-1}_{\overline{n}^q}(~i~) \ = \
      \kw{if} \ q=1 \ \kw{then} \ i \
      \kw{else} \ (\mathcal{B}^{-1}_{\overline{n}^{h=1\ldots q-1}}(i / n_{q}), \ i ~\%~ n_{q}) 
    \medskip\\
      \textbf{GenP}([\overline{n}^d],~f_{\overline{n}^d},~f^{inv}_{\overline{n}^d})\texttt{::apply}(\overline{i \ }^d) \ \ \ = \ \ f_{\overline{n}^d}(\overline{i \ }^d)
      \\
      \textbf{GenP}([\overline{n}^d],~f_{\overline{n}^d},~f^{inv}_{\overline{n}^d})\texttt{::inv}(~i_{flat}~) \ = \ f^{inv}_{\overline{n}^d}(~i_{flat}~)
      \\
      \textbf{GenP}([\overline{n}^d],~f_{\overline{n}^d},~f^{inv}_{\overline{n}^d})\texttt{::dims}() \ \ = \ \overline{n}^d
    \medskip\\
      \textbf{RegP}([\overline{n}^d],~\sigma_d)\texttt{::apply}(\overline{i \ }^d) \ \ = \ \ \mathcal{B}_{\sigma_d(\overline{n}^d)} (~\sigma_d(\overline{i \ }^d)~)
      \smallskip\\
      \textbf{RegP}([\overline{n}^d],~\sigma_d)\texttt{::inv}(~i_{flat}~) \ \ = \ \ \sigma_d^{-1}(~\mathcal{B}^{-1}_{\sigma_d(\overline{n}^d)}(~i_{flat})~)
      \\
      \textbf{RegP}([\overline{n}^d],~\sigma_d)\texttt{::dims}() \ \ = \ \overline{n}^d
    \medskip\\      
    \textbf{OrderBy}(\overline{Perm_d}^q)\texttt{::apply}(~\overline{i \ }^{d\cdot q}~) \ \ =
    \\\hsp 
      i_{flat} \leftarrow 0
    \\\hsp
      \kw{for} \ Perm \in \overline{Perm_d}^q ~\textbf{and}~k \in 0\ldots q-1 \ \kw{do}
    \\\hsp\hsp\hsp
        \overline{n}^d \ \leftarrow \ Perm\texttt{.dims}();
        \hsp\hsp
        \overline{i_{cur}}^d \leftarrow \overline{i \ }^{h=k\cdot d+1 \ldots k\cdot d + d}
    \\\hsp\hsp\hsp
        i^{cur}_{flat} \leftarrow Perm\texttt{.apply}(\overline{i_{cur}}^d)
    \\\hsp\hsp\hsp
        i_{flat} \leftarrow i^{cur}_{flat} \ + \ i_{flat}\cdot \prod_{h=1}^{d} (n_h)
    \\\hsp
      \kw{return} \ i_{flat}
    \\\smallskip
    \textbf{OrderBy}(\overline{Perm_d}^q)\texttt{::inv}(~i_{flat}~) \ \ =
    \\\hsp 
      \overline{i \ } \leftarrow \text{empty sequence}
    \\\hsp
      \kw{for} \ Perm \in \texttt{reverse}(\overline{Perm_d}^q) \ \kw{do}
    \\\hsp\hsp\hsp
        \overline{n}^d \ \leftarrow \ Perm\texttt{.dims}();
        \hsp\hsp
        p \leftarrow \prod_{h=1}^{d} (n_h)
    \\\hsp\hsp\hsp
        i^{cur}_{flat} \leftarrow i_{flat} \% p;
        \hsp\hsp\hsp\hsp
        i_{flat} \leftarrow i_{flat} / p
    \\\hsp\hsp\hsp
        \overline{i \ } \leftarrow Perm\texttt{.inv}(i^{cur}_{flat}), \ \overline{i \ }
    \\\hsp
      \kw{return} \ \overline{i \ }
    \\\smallskip
    \textbf{OrderBy}(\overline{Perm_d}^q)\texttt{::dims}(~) \ \ =
    \ \ 
      \overline{n} \leftarrow \text{empty sequence}
    \\\hsp
       \kw{for} \ Perm \in \overline{Perm_d}^q \ \kw{do}
    \hsp
       \overline{n} \leftarrow \overline{n}, \ Perm\texttt{.dims}()
    \\\hsp
    \kw{return} \ \overline{n}
    \end{array}
    \]
\end{small}
%
\vspace{-1ex}
\caption{Semantics of \texttt{apply} and \texttt{inv} of \textbf{OrderBy} Blocks.}
\label{fig:lego-order-by}
\end{figure}


LEGO's interface to the user consists of an \texttt{apply} and 
\texttt{inv} functions that can be called on a
\textbf{GroupBy} block:
\texttt{apply} receives as argument a multi-dimensional
index corresponding to the logical shape of \textbf{GroupBy},
and results in the corresponding flat index in the
(reordered) physical layout, while \texttt{inv} does the opposite. 
We define this functionality by a syntax-directed translation~\cite{torben-book},
detailed in Figures~\ref{fig:lego-order-by}~and~\ref{fig:lego-group-by},
which implements the \texttt{apply}, \texttt{inv} and \texttt{dims}
functions for each syntactic category of the LEGO language by 
combining the functionality of its syntactic constituents
(where \texttt{dims} is used to track the dimension sizes of a space). 

\textbf{GenP} simply applies the provided user-defined functions
$f$, $f^{inv}$. 
\textbf{RegP}'s \texttt{apply} flattens the index by applying the
canonical bijection $\mathcal{B}$ in the physical (permuted) layout,
hence the dimensions and index are permuted by $\sigma_d$.
Its \texttt{inv} unflattens the index by $\mathcal{B}^{-1}$ 
using the physical (permuted) dimensions and
recovers the logical-space index by permuting back
the physical index by the inverse~of~$\sigma_d$,
which is obtained by scattering $[1,\ldots,d]$ at 
the positions of $\sigma_d$.

\textbf{OrderBy}'s \texttt{apply} traverses the tiling 
space from outermost inwards, and at each steps flattens and 
accumulates the corresponding part of the index; \texttt{inv} 
unflattens the index from~innermost~outwards.

\begin{figure}
\begin{small}
  \hspace{-7ex}\vspace{-2ex}
    \[ 
    \begin{array}{l}
    \textbf{GroupBy}(~([\overline{n^1}^{d}],\ldots,[\overline{n^{q_g}}^d]), \ \overline{O}^v)\texttt{::apply}(~\overline{i \ }^{d\cdot q_g}~) \ \ =
    \\\hsp
      i_{flat} = \mathcal{B}_{(n^1_{1}\cdot\ldots\cdot n^{q_g}_d)}(~\overline{i \ }^{d\cdot q_g}~)
    \\\hsp
    \kw{for} \ O \ \in \ \texttt{reverse}(\overline{O}^v) \ \kw{do}
    \\\hsp\hsp\hsp
    \overline{n'^1}^{d'},\ldots,\overline{n'^{q_o}}^{d'} \leftarrow O\texttt{.dims}();
    \\\hsp\hsp\hsp
    \overline{i' \ }^{d'\cdot q_o} \leftarrow \mathcal{B}^{-1}_{n'^1_1\cdot\ldots\cdot n'^{q_o}_{d'}}(~i_{flat}~)
    \\\hsp\hsp\hsp
    i_{flat} \leftarrow O\texttt{.apply}(~\overline{i' \ }^{d'\cdot q_o}~)
    \\\hsp
    \kw{return} \ i_{flat}
    \smallskip\\
    \textbf{GroupBy}(~([\overline{n^1}^{d}],\ldots,[\overline{n^{q_g}}^d]), \ \overline{O}^v)\texttt{::inv}(~i_{flat}~) \ \ =
    \\\hsp
    \kw{for} \ O \ \in \ \overline{O}^v \ \kw{do}
    \\\hsp\hsp\hsp
      \overline{n'^1}^{d'},\ldots,\overline{n'^{q_o}}^{d'} \leftarrow O\texttt{.dims}();
    \\\hsp\hsp\hsp
      \overline{i' \ }^{d'\cdot q_o} \leftarrow O\texttt{.inv}(~i_{flat}~)
    \\\hsp\hsp\hsp
      i_{flat} \leftarrow \mathcal{B}_{n'^1_1\cdot\ldots\cdot n'^{q_o}_{d'}}(~\overline{i' \ }^{d'\cdot q_o}~)
    \\\hsp
    \kw{return} \ \mathcal{B}^{-1}_{(n^1_{1}\cdot\ldots\cdot n^{q_g}_d)}(~i_{flat}~)
    \smallskip\\
    \textbf{GroupBy}(~([\overline{n^1}^{d}],\ldots,[\overline{n^{q_g}}^d]), \ \overline{O}^v)\texttt{::dims}(~)
        = \overline{n^1}^{d},\ldots,\overline{n^{q_g}}^d
    \end{array}
    \]
\end{small}
\vspace{-1ex}
\caption{Semantics of \texttt{apply} and \texttt{inv} of \textbf{GroupBy} Blocks}
\label{fig:lego-group-by}
\end{figure}

Finally, Figure~\ref{fig:lego-group-by} shows the lowering
algorithm of \textbf{GroupBy}: its \texttt{apply} first flattens
its index in its logical space, and then traverses the chain
of reordering transformations $\overline{O}^v$ in reverse order, 
and for each one, denoted $O$, it remaps the flat index to $O$'s
logical space by $\mathcal{B}^{-1}$, and applies the reordering. 
%
\textbf{GroupBy}'s \texttt{inv} traverses $\overline{O}^v$
forward, and for each reorder $O$, it applies its inverse,
and then flattens it according to $O$'s logical space.
Ultimately, the resulting index is unflattened in 
\textbf{GroupBy}'s space.

\begin{figure}
\includegraphics[width=0.99\linewidth]{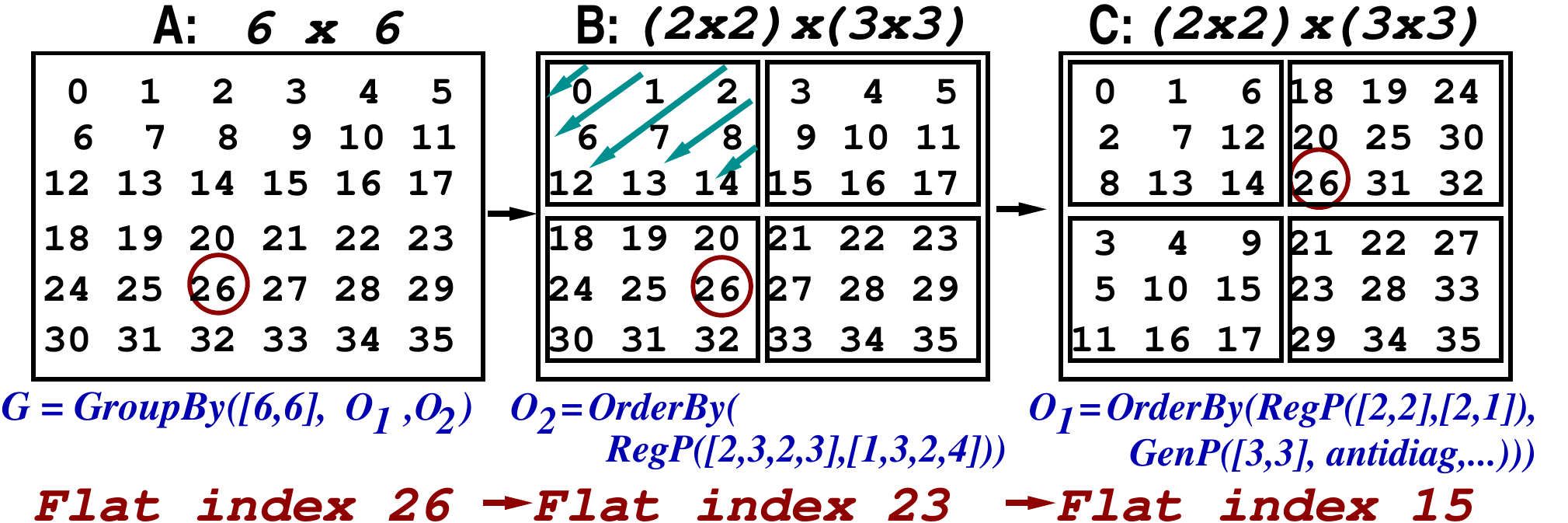}
\caption{
  $\mathbf{2\times 2\times 3\times 3}$ tiling 
  followed by transposing the outer dimensions and
  applying anti-diagonal permutation in the
  inner $\mathbf{3\times 3}$ blocks.
  The logical view is a $\mathbf{6 \times 6}$ matrix.
}
\label{fig:lego-complex-eg}
\end{figure}

Figure~\ref{fig:lego-complex-eg} demonstrates a more complex example
that uses a $6\times 6$ logical view, depicted in the left
column, whose elements correspond for convenience to the flat
index space $[0,\ldots,35]$. 

The middle column shows a reordering transformation $O_2$
that tiles the logical view into a $2\times 2$ grid
of $3\times 3$ blocks. This is achieved by a \textbf{RegP}
permutation that first stripmines each of the logical-view
dimensions of size $6$ into two smaller dimensions of sizes
$2\times 3$, and then interchanges the middle (second and third) 
dimensions, i.e.,
\begin{center}
\text{$O_2$ = \textbf{OrderBy}(\textbf{RegP}([2,3,2,3], $\sigma$=[1,3,2,4]))}
\end{center}

The right column of Figure~\ref{fig:lego-complex-eg} applies
another reordering $O_1$ that similarly
uses a hierarchical space of a $2\times 2$ grid of $3\times 3$
blocks, in which the grid is transposed (\textbf{RegP})
and the elements of each block are permuted (\textbf{GenP})
such that they are laid out in the order in which they appear
on the block's $2\cdot 3 - 1$ antidiagonals. In practice, the dimensions $d$ and $q$ are omitted, being implicitly inferred from the shape of the arguments:
\vspace{-3ex}

\begin{center}
\text{$O_1$ = \textbf{OrderBy}(\textbf{RegP}([2,2],$\sigma$=[2,1]), \textbf{GenP}([3,3], antidiag,..))}
\end{center}


\begin{figure}
  \begin{minipage}[t]{0.445\linewidth}
    \begin{lstlisting}[language=Python, basicstyle=\ttfamily\scriptsize,frame=none,mathescape=true]
def antidiag(n, i, j):
  antidg = i + j + 1
  if(antidg <= n):
    return i + (antidg*(antidg-1))/2
  else:
    antidg = 2*n - antidg
    gauss = (antidg * (antidg-1))/2
    return n*n - n + i - gauss
\end{lstlisting}
  \end{minipage}
  \begin{minipage}[t]{0.545\linewidth} 
    \begin{lstlisting}[language=Python, basicstyle=\ttfamily\scriptsize,frame=none,mathescape=true]
def antidiag$^{inv}$(n, x0):
  S = n*(n+1) / 2
  $x$ = x0 if x0 < S else n*n-1 - x0
  antidg = $\lfloor\sqrt{2 * x}\rfloor$
  antidg += ( $x$ >= (antidg*(antidg+1))/2 )
  i = $x$ - antidg*(antidg-1)/2
  j = antidg - i - 1
  return (i,j) if x0 < S else (n-1-i, n-1-j)
\end{lstlisting}
  \end{minipage}
  \vspace{-4ex}\\
\caption{Anti-Diagonal Permutation of an $n\times n$ Logical Space}
\label{fig:perm-egs}
\end{figure}

\noindent Finally, the pseudocode of the user-defined antidiagonal
permutation of a $n\times n$ logical space, and its inverse, 
are shown in Figure~\ref{fig:perm-egs}.
One can verify that the element at index $[4,2]$ in the $6\times 6$
logical view (left column), i.e., representing $26$, is reordered by
$O_2$ in the middle column to flat index $23$ (i.e., multi-dimensional
index [1,0,1,2]), and then by $O_1$ in the right column to physical
index $15$ (i.e., [0,1,2,0]). Conversely, one can use \texttt{inv} 
to compute that the flat physical index $15$ corresponds to
the logical-view index $[4,2]$.
\smallskip

\paragraph{\revised{Correctness}}
\revised{The algorithm presented in Figures~\ref{fig:lego-order-by}~and~\ref{fig:lego-group-by} provably implements a bijection, i.e., is correct by construction, if two assumptions hold: {\em First}, the user-defined function(s) $f^{inv}$ (and $f^{inv}_{\overline{n}^d}$) of $\textbf{GenP}([\overline{n}^d],~f_{\overline{n}^d},~f^{inv}_{\overline{n}^d})$ actually implement a bijection (and its inverse) from the multidimensional space $\mathcal{Z}_{n_1}\times\ldots\times \mathcal{Z}_{n_d}$ to the flat space $\mathcal{Z}_{n_1\cdot\ldots\cdot n_d}$. This is currently left as user's responsibility. {\em Second}, the total number of elements of the multidimensional spaces defined by the \textbf{GroupBy} and each of its contained \textbf{OrderBy} constructs is the same. This can be cheaply verified dynamically and even hoisted outside recurrences that do not affect these sizes. 
} 
\smallskip

\paragraph{\revised{Classification of Bijective Layouts}}
\revised{A single \textbf{OrderBy} construct using only regular permutations (\textbf{RegP}) implements an {\em affine} layout in the \texttt{apply} direction---i.e., a formula such as $i_!\cdot s_1 + \ldots + i_q \cdot s_q$, where $i_{1,\ldots, q}$ are the target indices and $s_{1,\ldots, q}$ are constant strides (a product of dimension~sizes). 

Sequencing two or more \textbf{OrderBy} constructs using only regular permutations {\em does not necessarily result in an affine layout} because the canonical bijections ($\mathcal{B}$ and $\mathcal{B}^{-1}$) applied at the borders introduce division and modulo operations that are not guaranteed to simplify away. 
Finally, user-defined permutations (\textbf{GenP}) allow {\em fully arbitrary} (bijective) layouts, e.g., using indirect arrays~\cite{bricks}, and quadratic (polynomial) indexing as in the anti-diagonal permutation of Figure~\ref{fig:perm-egs}.
}
\smallskip

\paragraph{\revised{Notation and Syntactic Sugar}} For convenience of presentation, this section has used the grammar in Figure~\ref{fig:lego-grammar}. The rest of the paper uses a notation that chains reordering and the final grouping transformations by means of dots:
%
\begin{equation}\label{layout-fig6}
\begin{array}{l}
\textbf{GroupBy}([6,6]).\\
\textbf{OrderBy}(\textbf{RegP}([2,3,2,3], [1,3,2,4])).\\
\textbf{OrderBy}(\textbf{RegP}([2,2], [2,1]),\\
    \hsp\hsp\hsp\hsp \textbf{GenP}([3,3], \text{antidiag}, \text{antidiag}^{inv}))
\end{array}
\end{equation}
%
\noindent As well, we define syntactic sugar for common operations:
\begin{small}
\[
\begin{array}{l}
\begin{array}{lcl}
\hspace{-1ex}\text{Row}([n_1,\ldots,n_d]) & \equiv & \textbf{RegP}([n_1,\ldots,n_d], \ [1,2,\ldots,d])\medskip\\
\text{Col}([n_1,\ldots,n_d]) & \equiv & \textbf{RegP}([n_d,\ldots,n_1], \ [d,\ldots,2,1])\medskip\\

\text{TileBy}_{q\times d}~( & \equiv &
       \textbf{GroupBy}([\overline{n^1}^d,\ldots,\overline{n^q}^d]).\\
\ [\overline{n^1}^d],\ldots,[\overline{n^q}^d]) &  & 
       \textbf{OrderBy}(\textbf{RegP}([\overline{n^1}^d,\ldots,\overline{n^q}^d], \sigma_{d\times q}))\medskip\\

\text{TileOrderBy}_{q\times d}& \equiv &
       \textbf{GroupBy}(P^1_d,\ldots,P^q_d).\\
\hsp(P^1_d,\ldots,P^q_d) & & \textbf{OrderBy}(~\textbf{RegP}(\\
  &  & \hspace{2ex}\sigma_{d\times q}(P^1_d.dims,..,P^q_d.dims, \sigma_{d\times q}^{-1}))\medskip\\
\end{array}\smallskip\\
\hspace{-1ex}\textbf{where} \ \sigma_{d\times q} = \text{flatten}(A), \textbf{with} ~ A:[d][q]\kw{int}, A_{k,h} = k+1+d\cdot h
\end{array}
\]
\end{small}


\text{Row} and \text{Col} define row- and column-major layouts,~corresponding to permuting dimensions by identity and by $[d,\ldots,1]$.
%
%
\text{TileBy$_{q\times d}$} denotes hierarchical tiling of $d$ dimensions on $q$ levels, e.g., 
\text{TileBy$_{3\times 2}$} and \text{TileBy$_{2\times 3}$} have 
permutations $\sigma_{2\times 3} = [1,3,5,2,4,6]$ and
$\sigma_{3\times 2} = [1,4,2,5,3,6]$, respectively, and applying
these permutations to their logical dimensions results, as expected, 
in the physical spaces
$(n^1_1\times n^2_1\times n^3_1)\times(n^1_2\times n^2_2\times n^3_2)$,
and
$(n^1_1\times n^2_1)\times(n^1_2\times n^2_2)\times(n^1_3\times n^2_3)$.
%
\text{TileOrderBy} similarly defines a hierarchical-tiling reordering.

\subsection{Comparison with CuTe/Graphene Algebra}
\label{subsec:comparison-graphene}

We identify two primary distinctions between the LEGO and CuTe/Graphene shape algebra representations.

\begin{table*}[htb]
\caption{Comparison of LEGO and CuTe/Graphene layouts for examples in figures and performance results.}
\label{tab:layout-comparison}
\begin{center}
\scriptsize
\begin{tabular}{|l|l|l|} \hline
\textbf{Fig.} & \textbf{LEGO} & \textbf{CuTe/Graphene}\\ \hline
&&\\ 
\ref{fig:lego_triton_motivation} &
$\text{TileBy}([M/BM, K/BK],[BM,BK]).\text{OrderBy}(\text{Row}(M,K))$ & $
(\begin{bmatrix}
M/BM & K /BK & BM & BK\\
K*BM & BK & K & 1
\end{bmatrix})$ \\[.1in]

\ref{fig:lego-complex-eg}mid & $\text{GroupBy}([6,6]).\text{OrderBy}(\text{RegP}([2,3,2,3],[1,3,2,4]))$ & 
$(\begin{bmatrix} 2, &2\\ 18, &3\end{bmatrix})\cdot(\begin{bmatrix} 3, &3\\ 6, &1\end{bmatrix})$\\[.15in] 
\ref{fig:lego_graphene_compare} &
$\text{GroupBy}([2, 2, 2, 2, 2] ).\text{OrderBy}(\text{RegP(} [2, 2, 2, 2, 2], [5, 2, 4, 3, 1]))$ &
$(\begin{bmatrix} 2, &2\\
1, &8\end{bmatrix})\cdot(\begin{bmatrix} 2, & (2, 2)\\ 2, & (4,16) \end{bmatrix})$\\[.15in]
\ref{fig:cuda-lud} & $\text{GroupBy}([R,R], [T,T]).\text{OrderBy}(\text{Row}(R * T, R * T))$ & $
(\begin{bmatrix}
R & R & T & T\\
 R T^2 & T^2 & T & 1
\end{bmatrix})$ \\[.1in]
\ref{fig:cuda-brick} &  $\text{TileBy}([N/B, N / B, N / B], [B, B, B]).\text{OrderBy}(\text{Row}(N/B, N / B, N / B), \text{Row}(B, B, B) )$  & $
(\begin{bmatrix}
N / B & N / B & N / B & B & B & B\\
N^2B & NB^2 & B^3 & B^2 & B & 1
\end{bmatrix})$\\[.1in] \hline
\end{tabular} 
\end{center}
\end{table*}

\paragraph{Elimination of Explicit Strides} 
LEGO supports all of the strided, rectangular layouts that can be expressed in the shape algebra for CuTe and Graphene.  A significant difference in the shape specification is that  
the CuTe/Graphene shape algebra requires the performance programmer to provide the strides for the layout, whereas LEGO derives the strides internally from the hierarchical tiling specification. Table~\ref{tab:layout-comparison} compares the LEGO and CuTE/Graphene layout specification for some of the various layouts used in the figures in this paper.  The simple tiled layout for the input matrices of Figure~\ref{fig:lego_triton_motivation} describes a 4D tiled data layout; the CuTe/Graphene layout in the third column linearizes the four dimensions to derive a stride.  
The tiled representation for CuTe/Graphene describing $G\circ O_2$ in Figure~\ref{fig:lego-complex-eg} 
%
 expresses the original layout $A$ of 6$\times$6 on the top row.  On the second row, the stride is specified: 6 between rows and 1 between columns within a row.  
 The resulting layout $B$ is (2$\times$2) $\times$ (3$\times$3), with a stride of 18 between block rows, and 3 between block columns.  The stride is 6 elements across tiles in the row dimension, and 1 in the column dimension.  

Even with this relatively simple tiled example, the need to specify strides already muddies the layout description.  However, the specification becomes more complex with the example of Figure~\ref{fig:lego_graphene_compare}, which matches Figure~4d in the Graphene paper~\cite{hagedorn2023graphene}.  In this case, as depicted in the figure, the goal is to create tiles (denoted by locations with the same color) that are not contiguous in either dimension.  In LEGO's formulation, 
this layout is simply a permutation of the five dimensions resulting from the tiling.  In contrast, Graphene expresses the layout with complex multi-dimensional strides. 

\begin{figure}[h]
\begin{center}
\begin{tikzpicture}[scale=0.6, transform shape]
\draw[step=0.75cm,black,thin] (0,0) grid (6,3);
\draw[very thick] (0,0) rectangle (1.5,0.75);
\draw[very thick] (0,1.5) rectangle (1.5,0.75);
\draw[very thick] (0,2.25) rectangle (1.5,1.5);
\draw[very thick] (0,3) rectangle (1.5,2.25);
\draw[very thick] (1.5,0) rectangle (3,0.75);
\draw[very thick] (1.5,1.5) rectangle (3,0.75);
\draw[very thick] (1.5,2.25) rectangle (3,1.5);
\draw[very thick] (1.5,3) rectangle (3,2.25);
\draw[very thick] (3,0) rectangle (4.5,0.75);
\draw[very thick] (3,1.5) rectangle (4.5,0.75);
\draw[very thick] (3,2.25) rectangle (4.5,1.5);
\draw[very thick] (3,3) rectangle (4.5,2.25);
\draw[very thick] (4.5,0) rectangle (6,0.75);
\draw[very thick] (4.5,1.5) rectangle (6,0.75);
\draw[very thick] (4.5,2.25) rectangle (6,1.5);
\draw[very thick] (4.5,3) rectangle (6,2.25);
\fill[cyan!20,semitransparent] (0,0) rectangle (1.5,0.75);
\fill[green!20,semitransparent] (0,1.5) rectangle (1.5,0.75);
\fill[cyan!20,semitransparent] (0,2.25) rectangle (1.5,1.5);
\fill[green!20,semitransparent] (0,3) rectangle (1.5,2.25);
\fill[pink!60,semitransparent] (1.5,0) rectangle (3,0.75);
\fill[blue!20,semitransparent] (1.5,1.5) rectangle (3,0.75);
\fill[pink!60,semitransparent] (1.5,2.25) rectangle (3,1.5);
\fill[blue!20,semitransparent] (1.5,3) rectangle (3,2.25);
\fill[cyan!20,semitransparent] (3,0) rectangle (4.5,0.75);
\fill[green!20,semitransparent] (3,1.5) rectangle (4.5,0.75);
\fill[cyan!20,semitransparent] (3,2.25) rectangle (4.5,1.5);
\fill[green!20,semitransparent] (3,3) rectangle (4.5,2.25);
\fill[pink!60,semitransparent] (4.5,0) rectangle (6,0.75);
\fill[blue!20,semitransparent] (4.5,1.5) rectangle (6,0.75);
\fill[pink!60,semitransparent] (4.5,2.25) rectangle (6,1.5);
\fill[blue!20,semitransparent] (4.5,3) rectangle (6,2.25);
\draw (0.375,2.625) node{\Large{0}};
\draw (1.125,2.625) node{\Large{4}};
\draw (1.875,2.625) node{\Large{8}};
\draw (2.625,2.625) node{\Large{12}};
\draw (3.375,2.625) node{\Large{16}};
\draw (4.125,2.625) node{\Large{20}};
\draw (4.875,2.625) node{\Large{24}};
\draw (5.625,2.625) node{\Large{28}};
\draw (0.375,1.875) node{\Large{1}};
\draw (1.125,1.875) node{\Large{5}};
\draw (1.875,1.875) node{\Large{9}};
\draw (2.625,1.875) node{\Large{13}};
\draw (3.375,1.875) node{\Large{17}};
\draw (4.125,1.875) node{\Large{21}};
\draw (4.875,1.875) node{\Large{25}};
\draw (5.625,1.875) node{\Large{29}};
\draw (0.375,1.125) node{\Large{2}};
\draw (1.125,1.125) node{\Large{6}};
\draw (1.875,1.125) node{\Large{10}};
\draw (2.625,1.125) node{\Large{14}};
\draw (3.375,1.125) node{\Large{18}};
\draw (4.125,1.125) node{\Large{22}};
\draw (4.875,1.125) node{\Large{26}};
\draw (5.625,1.125) node{\Large{30}};
\draw (0.375,0.375) node{\Large{3}};
\draw (1.125,0.375) node{\Large{7}};
\draw (1.875,0.375) node{\Large{11}};
\draw (2.625,0.375) node{\Large{15}};
\draw (3.375,0.375) node{\Large{19}};
\draw (4.125,0.375) node{\Large{23}};
\draw (4.875,0.375) node{\Large{27}};
\draw (5.625,0.375) node{\Large{31}};
\end{tikzpicture}
\end{center}
\vspace{-1ex}
\caption{Example layout that is non-contiguous in 2 dimensions: LEGO and Graphene layout specifications shown in Table~\ref{tab:layout-comparison}.}
\label{fig:lego_graphene_compare}
\end{figure}

\paragraph{Extended Layout Support} LEGO is not limited to strided layouts; it also accommodates additional layouts that require complex indexing expressions beyond rectangular, strided layouts. For example, 
the antidiagonal layout for $O_1$ in Figure~\ref{fig:lego-complex-eg}, whose implementation is described in Figure~\ref{fig:perm-egs} and Equation~\ref{layout-fig6}, cannot be supported by the CuTe/Graphene shape algebra.  Because LEGO can represent any bijective mapping between physical and logical layout, it can represent this antidiagonal, and, as discussed in Section~\ref{sec:concl}, provides a foundation for other
commonly-used bijective layouts. 

\subsection{Beyond Bijective Layouts}

\revised{
In some cases, LEGO primitives can be composed to support certain layouts that are not bijective. \amirbijective{Importantly, we support partial tiles where the tile size does not evenly divide the problem size, adopting a similar approach to the oversampling method in CuTe~\cite{CuTE}} as follows.
A new constructor \textbf{ExpandBy}, as illustrated in Figure~\ref{fig:lego-expand-by}, performs the necessary widening/narrowing conversions between a physical $d$-dimensional space $\overline{n}^{\,d}$ whose sizes do not evenly divide the tiles, and an extended one $\overline{n'}^{\,d}$ that does, such that the bijective layout $G$ is safely applied in the expanded space. 
%
Specifically, \texttt{apply} projects a logical index through \textit{G} to a flat index in the expanded layout, unfolds it via the canonical bijection $\mathcal{B}$, accepts it only if the coordinates fall within the original physical extents, and reports the corresponding flat position in the original space (otherwise $-1$). Conversely, \texttt{inv} lifts to an original flat index, re-flattens it in the expanded space via the canonical bijections, and then inverts through \textit{G}.
}

\revised{

To accommodate injective layouts such as broadcasting \((i,j)\mapsto i\) or \((i,j)\mapsto j\) and even-mapping \(i\mapsto 2i\), we restrict the language to exporting only \texttt{apply} (not \texttt{inv}) and to using exactly one \textbf{GroupBy} followed by an \textbf{OrderBy} of the same shape, where that \textbf{OrderBy} contains a single \textbf{GenP} that may be injective.
For the remainder of the paper, we focus on bijective layouts (i.e., \textbf{GenP} being bijective).
}

%


\begin{figure}[t] 
\revised{
\begin{flalign*}
&
\begin{array}{@{}l@{}}
ExpandBy \ ::= \ \textbf{ExpandBy}(\overline{Tile_{d}}, \ \overline{Tile_{d}}, \ GroupBy)
\smallskip\\
%
\textbf{ExpandBy}([\overline{n}^{\,d}],[\overline{n'}^{\,d}],G)\texttt{::apply}(~\overline{s}^{\,q}~)\ \ =\\
\hsp i_{\mathrm{flat}} \leftarrow G\texttt{::apply}(~\overline{s}^{\,q}~)\\
\hsp \overline{i'}^{\,d} \leftarrow \mathcal{B}^{-1}_{\;n'_1\cdot\ldots\cdot n'_d}(~i_{\mathrm{flat}}~)\\
\hsp \kw{if}\ (i'_1<n_1)\wedge\cdots\wedge(i'_d<n_d)\ \kw{then}\\
\hsp\hsp \kw{return}\ \mathcal{B}_{\;n_1\cdot\ldots\cdot n_d}(~\overline{i'}^{\,d}~)\\
\hsp \kw{else}\ \kw{return}\ -1
\smallskip\\
\textbf{ExpandBy}([\overline{n}^{\,d}],[\overline{n'}^{\,d}],G)\texttt{::inv}(~i_{\mathrm{flat}}~)\ \ =\\
\hsp \overline{i}^{\,d} \leftarrow \mathcal{B}^{-1}_{\;n_1\cdot\ldots\cdot n_d}(~i_{\mathrm{flat}}~)\\
\hsp i'_{\mathrm{flat}} \leftarrow \mathcal{B}_{\;n'_1\cdot\ldots\cdot n'_d}(~\overline{i}^{\,d}~)\\
\hsp \kw{return}\ G\texttt{::inv}(~i'_{\mathrm{flat}}~)
\smallskip
\end{array}
&&
\end{flalign*}
}
\vspace{-3ex}
\caption{ Grammar and (\texttt{apply}/\texttt{inv}) semantics of \textbf{ExpandBy}.
}
\label{fig:lego-expand-by}
\end{figure}

\section{Integrating LEGO into Ecosystems}
\label{sec:Implementation}
As demonstrated in previous sections, LEGO establishes an algebraic framework independent of any compilation system.
We see it as an important tool that can be integrated into a compiler or code generator, particularly \revised{to support data/thread tile abstractions} for GPUs, but also applicable to threaded CPU code generation and future heterogeneous hardware.  To demonstrate the power of the LEGO indexing mapping from layout specification to code generation, it was essential to integrate LEGO into mature ecosystems and rely on these to optimize code resulting from the layout mapping. 

In this section, we describe the integration of LEGO into Triton, CUDA, and MLIR. The integration with Triton and CUDA illustrates a straightforward implementation using Python. The incorporation within MLIR underscores the tool's versatility.  

\subsection{Code Generation via Instantiating Templates}

Our implementation for generating Triton and CUDA code utilizes 
an approach in which the user supplies code containing placeholders, and separately-defined layouts.
 The placeholders, marked using the Jinja2~\cite{jinja2} syntax \texttt{\{\{~\}\}}, are intended to represent index expressions or layout-specific logic.   LEGO then generates appropriate symbolic expressions based on the user-defined layout and replaces the corresponding placeholders within the template. This process offloads the complexity of constructing low-level index calculations from the user.
 An example of this  specification 
 was shown in Figure~\ref{fig:lego_triton_motivation}
(right).  
 
For this purpose, the LEGO algebra is integrated into the SymPy framework~\cite{sympy}, a Python library for symbolic mathematics. This integration enables advanced symbolic reasoning and high-level manipulation of index expressions, including algebraic simplification.
However, SymPy does not have all the necessary information to generate the optimized index expression. In particular, it lacks details about the range of variables used to index into the layout. We propagate this range information through the layout and develop a custom SymPy expression traversal that leverages these range constraints to simplify the index expressions. 
Moreover, since our algebra involves modulo and floor-division operations, we apply seven custom simplifications summarized in Table~\ref{tab:simplification_rules_small}. Each rule’s side-conditions (e.g.\ non-negativity and upper-bound checks) are proved by the Z3 SMT solver~\cite{de2008z3} using the index ranges derived from the layout specification. In addition, users can provide their own constraints to the system to further simplify the expression.
The indexing code is then generated by the Python and C printers provided by SymPy. 

\begin{figure}[!ht]
\centering
    \includegraphics[width=1\linewidth]{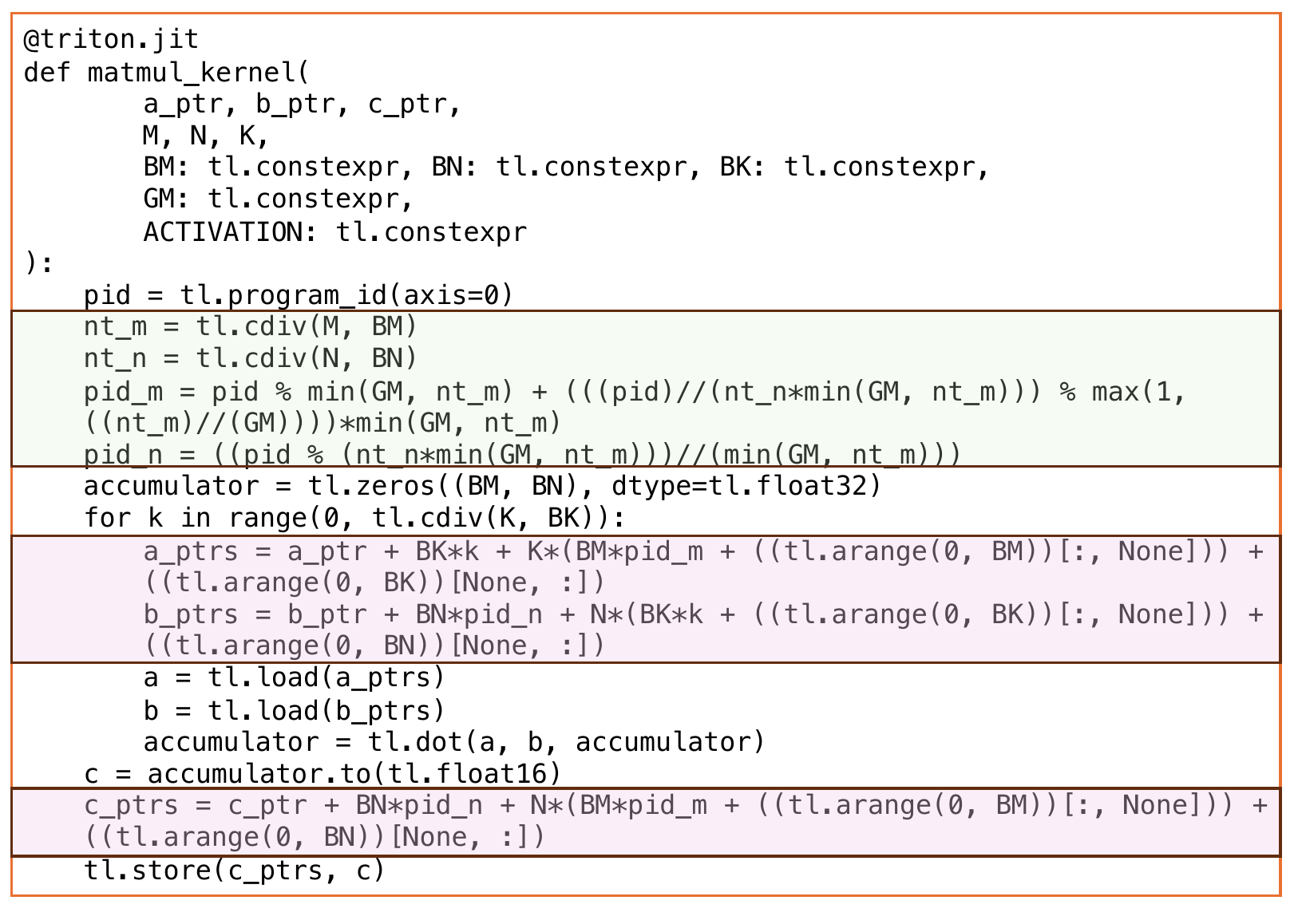}
\caption{LEGO layouts instantiated into  Triton template of Figure~\ref{fig:lego_triton_motivation}~(right).}
\label{fig:triton_output}
\vspace*{-1ex}
\end{figure}

For integration with Triton, we have introduced specialized slicing syntax analogous to NumPy's slice notation~\cite{numpy}. Specifically, when a user employs a colon (\texttt{:}) to denote the entire dimension—specified through \text{TileBy}—the system generates a corresponding \texttt{tl.arange}, whose bounds are derived from the layout specifications. Furthermore, Triton mandates that the upper and lower bounds of this range be known at the time of compilation. We show the final Triton code generated by this process in Figure~\ref{fig:triton_output}, starting with the input from Figure~\ref{fig:lego_triton_motivation}
(right).


\begin{table}[ht]
\centering
\caption{Integer division and modulo simplification rules.}
\label{tab:simplification_rules_small}
\begin{tabular}{|p{3cm}|p{2cm}|p{2.4cm}|}
\hline
\textbf{Pattern} & \textbf{Result} & \textbf{Condition} \\
\hline
\texttt{(d*q + r) \% d} 
& \texttt{r \% d} 
& $d \ne 0$ \\
\hline
\texttt{(d*q + r) / d} 
& \texttt{q} \newline \texttt{q + r / d} 
& $d \ne 0,\; 0 \le r < d$ \newline otherwise \\
\hline
\texttt{(x \% d) / d} 
& \texttt{0} 
& $d > 0$ \\
\hline
\texttt{x / a} 
& \texttt{0} 
& $a > 0,\; 0 \le x < a$ \\
\hline
\texttt{x \% a} 
& \texttt{x} 
& $a > 0,\; 0 \le x < a$ \\
\hline
\texttt{(n + y) / 1} 
& \texttt{n + (y / 1)} 
& $n \in \mathbb{Z}$ \\
\hline
\texttt{a*(x / a) + x \% a} 
& \texttt{x} 
& $a \ne 0$ \\
\hline
\end{tabular}
\end{table}

%



We also evaluated whether pre-expanding terms in index expressions before invoking SymPy’s simplification routines (including our range-simplification pass) improves performance compared with simplifying the unexpanded expressions, since expansion can reveal additional optimization opportunities. In the NW benchmark, skipping pre-expansion produced better performance by minimizing the total number of operations. In the LUD benchmark, pre-expansion helped by exposing simplifications that lowered the operation count and improved runtime. To accommodate both cases, we use a simple cost model that counts operations in the generated expression and selects the variant with the lowest count, choosing the unexpanded form for NW and the expanded form for LUD.



\subsection{End-to-end Code Generation in MLIR}

MLIR facilitates end-to-end code generation through its robust dialect system. We integrate LEGO into MLIR by using the previously-described SymPy expression simplification, creating a custom SymPy printer using the MLIR Python bindings. In this framework, the layout algebra is implemented with the \texttt{arith} and \texttt{affine} dialects for arithmetic and control flow operations, the \texttt{memref} and \texttt{vector} dialects for managing memory operations, and the \texttt{gpu} dialect providing GPU code generation primitives. This approach leverages Python bindings to ensure compatibility with both standard MLIR dialects and custom user dialects, which can take advantage of LEGO layout algebra for their specialized implementations. A single MLIR file is then generated, encapsulating both layout information and compute code. 


By integrating LEGO into MLIR, there is the potential for broader adoption in domain-specific frameworks, including but not limited to tensor computations. We demonstrate this implementation here, but such an integration with other dialects will be the subject of future work.

\section{Evaluation}
\label{sec:results}

The goal of the performance evaluation is to demonstrate LEGO data and thread-block layouts support code generation in a variety of contexts, and integrated with 
various state-of-the-art frameworks.  

\begin{enumerate}


\item \textit{Triton to demonstrate integration in state-of-the-art DSLs}, e.g., targeting tensor-core utilization. 
We match Triton's performance while simplifying the input specification.  

\item \textit{CUDA to demonstrate integration in a mainstream GPU programming language}, and the benefits of a richer set of data and thread-block layouts. 

\item \textit{MLIR to demonstrate integration in mainstream compiler frameworks}. 
\end{enumerate}

LEGO data layouts used in the experiments include 2D tiles that are in row-major or column-major order. 
We also employ a novel 3D brick data layout in the CUDA experiments. For thread layouts, 2D thread blocks in row major order are common, but we also use 3D thread blocks for bricks, an antidiagonal for NW, and a thread coarsening pattern for LUD (please refer to Table~\ref{tab:layout-comparison}).

\revised{
Table~\ref{tab:gen-times} summarizes the one-time code-generation and simplification latency for each application on a personal laptop (Apple M2 Max). The generation and simplification time of the per-application code ranges from sub-second to several seconds. 
This overhead is limited to index-generation time and does not affect steady-state execution.

\begin{table}[htb]
  \centering
  \caption{\revised{Per-application code generation and simplification.}}
  \label{tab:gen-times}
  \revised{

  \begin{tabular}{|p{5cm}|p{2cm}|}
    \hline
    \textbf{Benchmark} & \textbf{Generation time} \\
    \hline
    Layernorm FWD + BWD      & 0.33 s \\
    \hline
    Grouped GEMM             & 0.65 s \\
    \hline
    Softmax                  & 0.05 s \\
    \hline
    Matmul (each variant)    & 1.11 s \\
    \hline
    LUD                      & 0.87 s \\
    \hline
    NW                       & 0.46 s \\
    \hline
    Bricks (Cube/Star)       & 5.95 / 18.07 s \\
    \hline
    Transpose (Naive/SMEM)   & 1.07 / 1.15 s \\
    \hline
  \end{tabular}
    }
\end{table}

}
In this study, the experiments were executed using an NVIDIA Ampere A100 80GB GPU, deployed on an AMD EPYC 7513 processor with 32 cores under a CentOS operating system. The experimental framework was configured with LLVM (commit 556ec4a), Triton 3.2.0, PyTorch 2.5.1, and CUDA 12.4. Each benchmark was executed 25 times for warm-up, followed by 100 repetitions for data collection, and the mean performance value from these repetitions is reported. 

\begin{table}[htb]
\centering
\caption{\revised{Arithmetic ops before and after optimization.}}
  \revised{
\begin{tabular}{|p{3cm}|p{2cm}|p{2cm}|}
\hline
\textbf{Operator} & \textbf{Original Ops} & \textbf{Optimized Ops} \\
\hline
LayerNorm (FWD)  & 6  & 1 \\
\hline
LayerNorm (BWD)  & 4  & 0 \\
\hline
Softmax          & 4  & 0 \\
\hline
Grouped GEMM     & 20 & 6 \\
\hline
Matmul           & 31 & 9 \\
\hline
\end{tabular}
}
\label{tab:ops_reduction}
\end{table}

\subsection{Triton Benchmarks}
\begin{figure*}[t]
  \centering
  \begin{subfigure}[t]{0.32\textwidth}
    \centering
    \includegraphics[width=\linewidth]{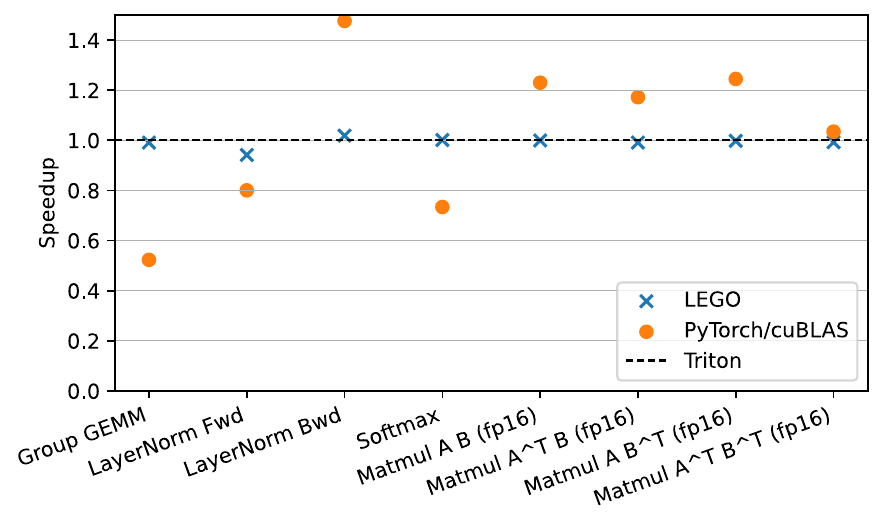}
    \caption{$N=2048$}
    \label{fig:speedup-2048}
  \end{subfigure}\hfill
  \begin{subfigure}[t]{0.32\textwidth}
    \centering
    \includegraphics[width=\linewidth]{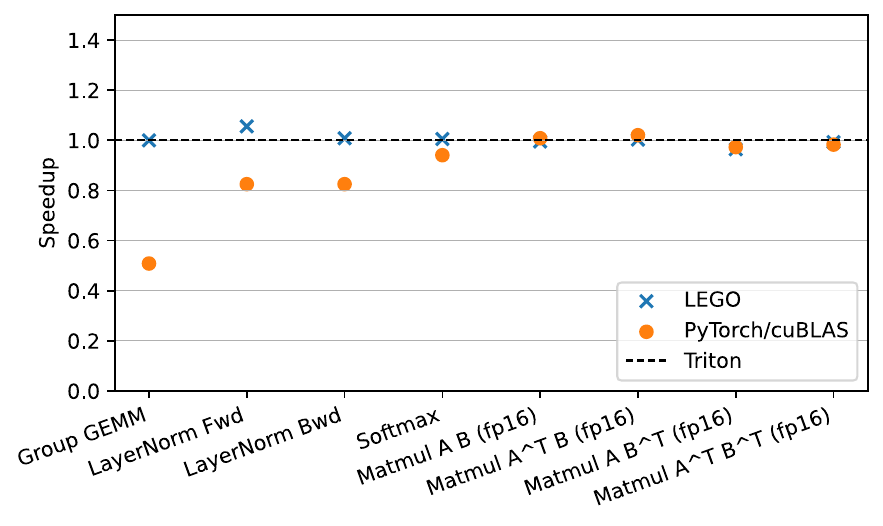}
    \caption{$N=4096$}
    \label{fig:speedup-4096}
  \end{subfigure}\hfill
  \begin{subfigure}[t]{0.32\textwidth}
    \centering
    \includegraphics[width=\linewidth]{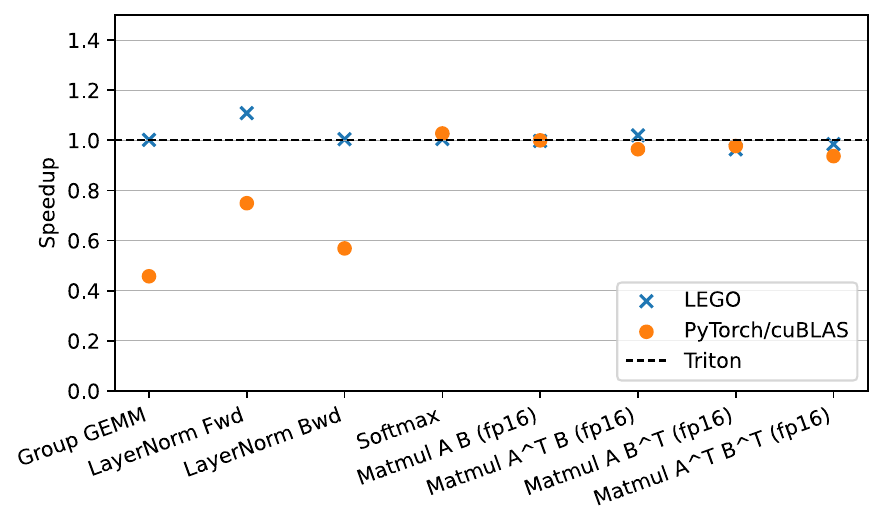}
    \caption{$N=8192$}
    \label{fig:speedup-8192}
  \end{subfigure}
  \caption{Performance comparison of Triton and PyTorch against the generated code using LEGO.}
  \label{fig:eval_triton}
\end{figure*}

In this study, we evaluated the LEGO framework using five benchmarks obtained from the official Triton repository: group GEMM, LayerNorm Fwd, LayerNorm Bwd, softmax, and matrix multiplication in FP16. These benchmarks were selected due to their computational heterogeneity and their frequent application in machine learning workloads. Performance of LEGO versions was measured for three problem sizes 
against reference implementations from the Triton repository, as shown in Figure~\ref{fig:eval_triton}.
\revised{We use Triton as our baseline. PyTorch’s CUDA backend dispatches matrix multiplications to cuBLAS, resulting in vendor-optimized kernels.}

Overall performance is nearly identical between the LEGO and Triton benchmarks.  PyTorch/cuBLAS outperforms both for most benchmarks at 2k size, but as the problem size gets larger, the LEGO-generated code is able to fully utilize the tensor cores.  
For matrix multiplication experiments, we used power-of-two square matrices. 
We selected configurations that avoided partial tiling in the inputs, thereby eliminating the need for load/store masking in the Triton kernel, ensuring a fair comparison.
Four variations of matrix multiplication were generated 
using the generic kernel template discussed in the previous section, with the only modification being the data layout for matrices \(A\) and \(B\). The transposed version employs a column-major layout (\(Col\)), while the non-transposed version utilizes a row-major layout (\(Row\)); for example, in the case of \(AB^T\), where \(A\) is \(Row(M, K)\) and \(B\) is \(Col(K, N)\). This highlights the flexibility of LEGO code generation, demonstrating that by merely altering the data layout, different implementations of matrix multiplication can be achieved.
As illustrated in Figure~\ref{fig:eval_triton}, LEGO achieves performance on matrix multiplication comparable to that of Triton and PyTorch/cuBLAS.

For the remaining benchmarks, LEGO and Triton outperform PyTorch/cuBLAS in some cases, although the difference is small with softmax.
LEGO also outperforms Triton on the LayerNorm Fwd benchmark because the example in the original repository uses a for loop with an explicit step, which Triton’s codegen handles less efficiently than loops with a manually incremented step. For LayerNorm Bwd, we benchmark only the backward pass and skip the forward pass.

\revised{
For program specification, arithmetic operations in user-defined code were reduced across evaluated components (Table~\ref{tab:ops_reduction}), demonstrating LEGO’s ability to generate high-performance Triton kernels with 
simpler expressions
}

\begin{figure*}[h]
  \centering
  \begin{subfigure}[t]{0.32\textwidth}
    \centering
    \includegraphics[width=\linewidth]{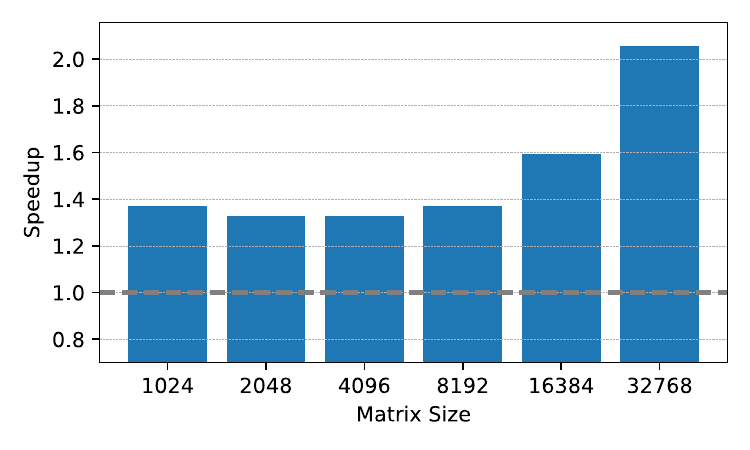}
    \caption{NW}
    \label{fig:cuda-nw}
  \end{subfigure}\hfill
  \begin{subfigure}[t]{0.32\textwidth}
    \centering
    \includegraphics[width=\linewidth]{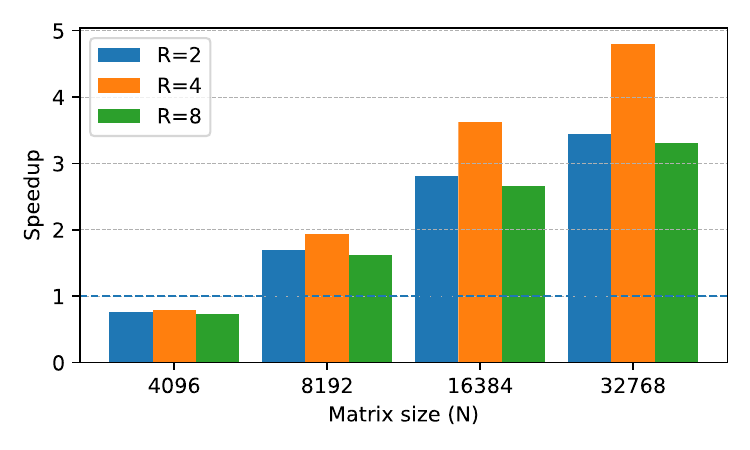}
    \caption{LUD}
    \label{fig:cuda-lud}
  \end{subfigure}\hfill
  \begin{subfigure}[t]{0.32\textwidth}
    \centering
    \includegraphics[width=\linewidth]{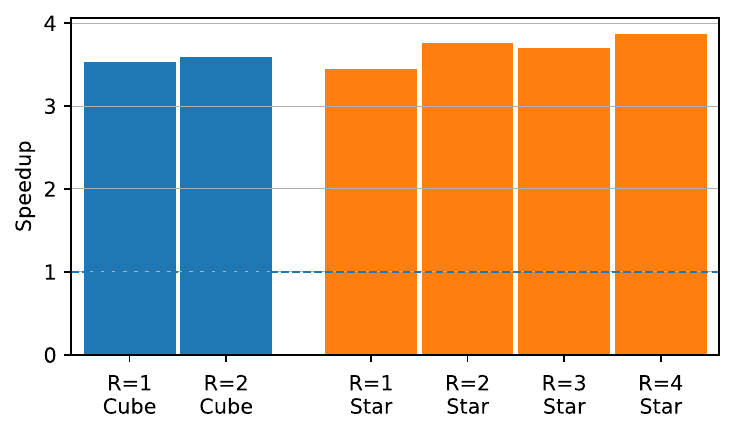}
    \caption{Stencil}
    \label{fig:cuda-brick}
  \end{subfigure}\hfill
  \caption{Performance comparison of CUDA benchmarks against Rodinia benchmark for NW and LUD. We compare brick vs. array layout for stencil examples.}
  \label{fig:cuda_perf}
\end{figure*}

\subsection{CUDA Benchmarks}

To illustrate layouts not supported by the other frameworks and the efficacy of exploring different data or thread-block layouts, we present three examples in Figure~\ref{fig:cuda_perf}.  

The first is the NW benchmark from the Rodinia benchmark suite~\cite{rodinia}.   Its CUDA implementation consists of two kernels that are called in a loop executed on the host. The kernels utilize a $(b+1)\times(b+1)$ buffer \texttt{buff} that is maintained in shared memory, and whose elements on each anti-diagonal are updated in parallel. Since Rodinia requires $b$ to be a multiple of $16$ and $b$ is also the size of the CUDA block, it follows that the read and write accesses of the original code exhibit stride $b$, resulting in expensive bank conflicts. 

We optimize \texttt{buff}'s layout by applying the anti-diagonal reordering (permutation) shown in Figure~\ref{fig:perm-egs}, which uses the indexing expressions generated by LEGO, and by overloading the \texttt{[]} operator to redirect logical accesses from the original code. This requires the definition of a small wrapper class for arrays and the modification of only two lines of the original code.  LEGO's layout description is shown in Equation~\ref{layout-fig6} of Section~\ref{subsec:lego-blocks}.
%
%
%
%
As demonstrated in Figure~\ref{fig:cuda-nw}, this layout transformation improves performance from  \(1.4\times\) up to \(2.1\times\) \revised{by reducing shared memory bank conflict and warp stalls}.

\begin{figure}[htb]
  \centering
  \begin{subfigure}{0.49\linewidth}
    \centering
    \includegraphics[width=\linewidth]{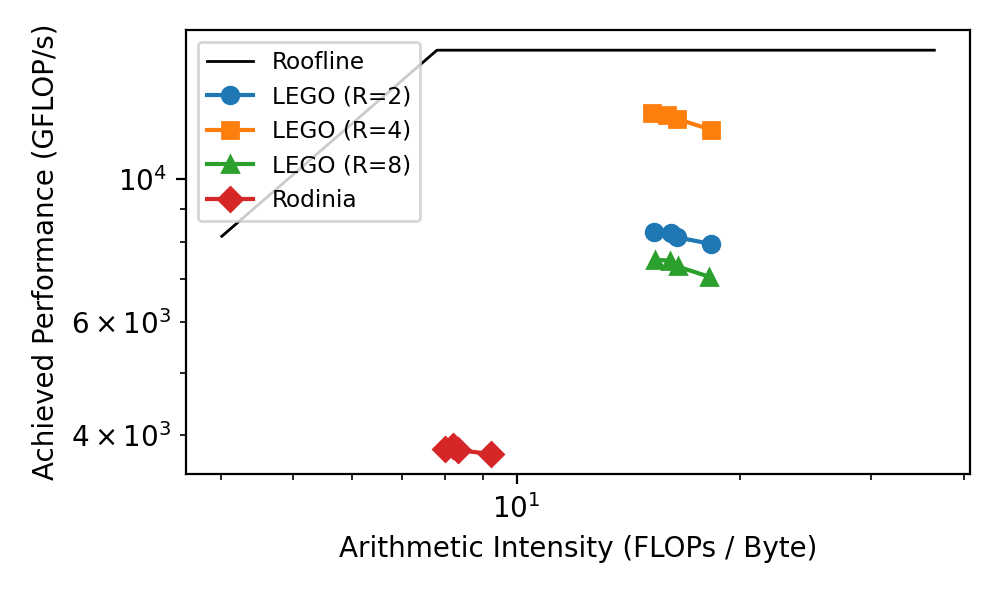}
    \caption{LUD}
    \label{fig:cuda-lud-roofline}
  \end{subfigure}
  \hfill
  \begin{subfigure}{0.49\linewidth}
    \centering
    \includegraphics[width=\linewidth]{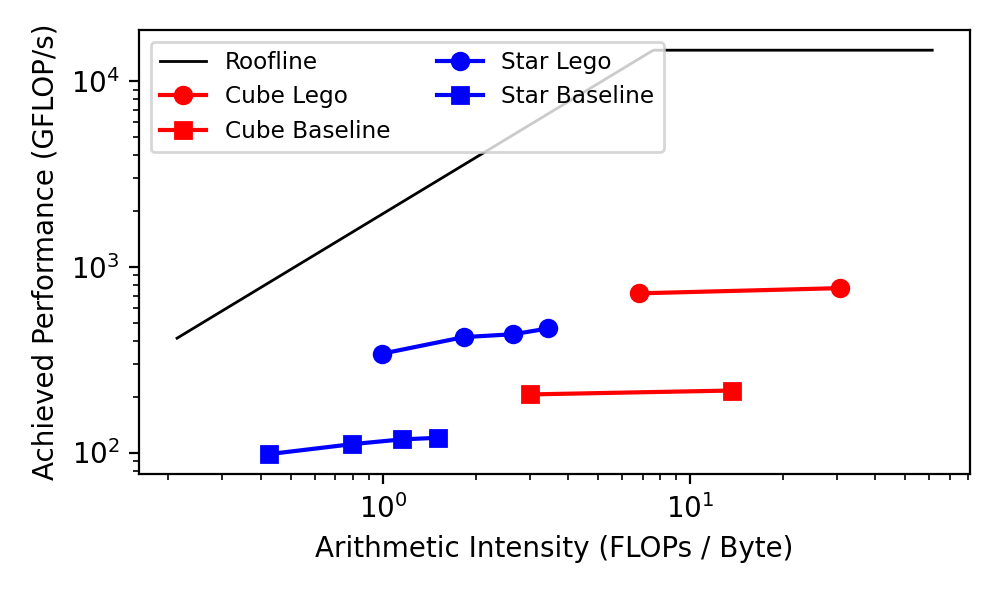}
    \caption{Stencil}
    \label{fig:cuda-stencil-roofline}
  \end{subfigure}
  \caption{\revised{Roofline performance for LUD \& Stencil benchmarks.}} 
  \label{fig:cuda-roofline}
\end{figure}

The second CUDA example is LUD, also from the Rodinia benchmark suite.  For this example, we apply a common optimization called thread coarsening, whereby the amount of work performed by each thread is increased~\cite{Barua18,Ivanov24}.  But what distinguishes our approach is that thread coarsening is re-imagined as a layout optimization.  LEGO's thread-block layout optimization binds values to both the total number of threads in each dimension and the bounds on the outer loop in the thread.  The layout description is provided in Table~\ref{tab:layout-comparison}, and the resulting performance \revised{and roofline are shown in Figure~\ref{fig:cuda-lud} and Figure~\ref{fig:cuda-lud-roofline}}. Although the baseline code uses a logical LUD block size of $16\times 16$ and a one-to-one correspondence to the CUDA block size, the best performance is obtained with an LUD block size of $64\times 64$ and a coarsening factor of 4, which keeps the CUDA block size at $16\times 16$ yet executes more work per thread block \revised{and achieves best block-level parallelism}.

The final example illustrates a modified data layout for 3D stencil computations.
The benchmark is based on the array and 
brick data layout code in Zhou et al.~\cite{brick-p3hpc18}.  Bricks are 3D subdomains stored in contiguous memory, so that spatially adjacent data related to a block of computation are also physically adjacent, thus eliminating unnecessary data movement over strided data when a conventional row-major layout is used~\cite{bricks}.
The brick layout 
-- in the last row of Table~\ref{tab:layout-comparison} -- 
is a 6D object; the CuTe/Graphene layout expresses a stride for each dimension.
The experiment compares a row-major layout to a brick layout for 3D cube-shaped (27-pt and 125-pt) and star-shaped (7-pt, 13-pt, 19-pt, and 27-pt) stencils. \revised{As shown in Figures~\ref{fig:cuda-brick} and~\ref{fig:cuda-stencil-roofline}, we observe speedups of $3.4\times$–$3.9\times$ across all stencil types solely from changing the data layout, even without integration with vector code generation
described in Zhou et al.~\cite{bricks}}.

\subsection{MLIR Benchmark}

To demonstrate an integration into a compiler framework, we evaluate LEGO's MLIR GPU code with a 2D transpose operation, 
a simple example to showcase optimization of data movement.  
Table~\ref{tab:MLIR_results} shows a comparison between the LEGO-MLIR implementation, compiled from MLIR, and the baseline code from the NVIDIA CUDA, compiled with \texttt{nvcc}.  In the \textbf{Naive} code, the input and output matrices are read/written from global memory, resulting in uncoalesced global memory accesses. \textbf{Smem+Coalesced} stages data in shared memory (another layout in LEGO) so that all global memory accesses are coalesced.  Results are listed as throughput (GB/s).  
In spite of using different compilation frameworks, performance results are comparable, with a slight advantage to LEGO for generating linearized array accesses.

\section{Related Work}
Deriving high-performance implementations of tensor computations is a fertile area of active research.  We will focus this section on a narrow set of prior work that integrates data layout and/or data movement into code generation.  

\begin{table}[htb]
\caption{Comparison of LEGO performance on 2D transpose with CUDA SDK baseline using MLIR. Performance is reported in GB/s throughput, so higher numbers are better.} \label{tab:MLIR_results}
\begin{tabular}{|l|r|r|r||r|r|r|} \hline
& \multicolumn{3}{|c||}{\textbf{Naive}} & \multicolumn{3}{c|}{\textbf{Smem+Coalesced}}\\ \hline
\textbf{Size}& \textit{2048} & \textit{4096} & \textit{8192} & \textit{2048} & \textit{4096} & \textit{8192}\\ 
\hline \hline

\textbf{CUDA-SDK} & 212.0 & 175.8 & 175.4& 670.0 & 718.2 & 735.7  \\ \hline
\textbf{LEGO-MLIR} & 206.8 & 178.0 & 190.7 & 681.7 & 741.2 & 759.4\\ \hline
\end{tabular}

\end{table}

\paragraph{Data movement specifications}
Historically, data copy was applied in compilers to reorganize submatrices, especially to avoid conflict misses in cache or stage data in explicitly managed storage~\cite{Temam93,Saday08}. 
CUDA-CHiLL incorporated datacopy into its scheduling language to copy data to/from global memory, shared memory, and texture memory in GPUs~\cite{CUDA-CHiLL}.  More recently, Fireiron and MDH enrich these data movement specifications for GPUs~\cite{hagedorn2020fireiron,10.1145/3578360.3580269}.

\paragraph{Data layout 
for sparse tensors}
Specifying data layout is central to optimizing sparse matrix and tensor computations, where the representation of only nonzero elements varies to better exploit their structure.
Moreover, loop optimizations must be reformulated whenever loop indices iterate over a sparse dimension of a tensor~\cite{STROUT201632, venkat:cgo14,venkat:pldi15,venkat:sc16}.  TACO~\cite{kjolstad2017tensor} introduced an approach to co-iteration over multiple sparse tensors, where the intersection (for multiply) or the union (for addition) of the nonzero locations must be identified.  The user specifies the layout along with the computation in Einstein notation, and the compiler generates the code for the input with the specified layout. To improve the performance and take advantage of the optimized data layouts, code transformations were later enabled in {TACO} through a scheduling language~\cite{senanayake2020sparse}. 
Where logical indices may not have corresponding physical entries, the inverse mapping from physical to logical indices can be used to find corresponding elements in other tensors during co-iteration, as done in dlcomp~\cite{dlcomp}.  

\paragraph{Data layout 
for performance portability}
Data layouts such as Kokkos View~\citep{kokkosview2024}, and \texttt{std::mdspan} in C++ 2023, abstract away the underlying data organization in memory for performance portability.
Their underlying data layouts exploit the hierarchical nature of GPUs and CPU/GPU systems.  In structured grid computations, fine-grained data blocking, where logically adjacent three-dimensional subdomains are stored in contiguous memory, have been shown to significantly reduce data movement~\cite{Araya-Polo:2009:SIT:1507443.1507449,bricks,yount2016yask}.  TiDA~\citep{Unat2016,LocalityAbstractions} uses coarse-grained data blocking, where the entire grid is tiled into sub-grids, each with its own ghost zone.   

\paragraph{Data layout and 
thread layout applied to tensors}
As previously noted, LEGO is most closely connected to the approaches of Graphene~\cite{hagedorn2023graphene} and Triton~\cite{triton,triton2}, which are focused on utilizing tensor cores.  Graphene uses the same layout specification for data layout/movement and thread/block layout, representing general strided rectangular regions, as in Figure~\ref{fig:lego_graphene_compare}.  As shown in Figure~\ref{fig:lego_triton_motivation}, Triton's layout specification provides a slice of each tensor, but requires explicit stride calculations.  
Very recently, linear layouts have been introduced; these are internal to the Triton compiler and not exposed at the source code level~\cite{triton2}.  The layout description eliminates the need for explicit strides, but instead uses a linear algebra formulation where 
a layout is described with a binary matrix.
Moreover, linear layout does not support user-defined bijective layouts that are nonlinear.
At the application level, a tensor library called einops~\cite{einops} exposes a tensor notation to describe tensor structure, facilitating the integration of tensor libraries. 






\paragraph{Distributed Data Layouts}
Common array layouts, such as tiled, row- and column-major, have been supported for a long time as directives in languages for distributed programming, such as High-Performance Fortran~\cite{HP-Fortran}. ZPL~\cite{ZPL} separates the definition of the hardware abstraction from the manner in which data is mapped to the hardware,
and Sequoia~\cite{Sequoia} and Legion~\cite{Legion} build on this idea to support, for example, 
(1) hierarchical definition of the hardware,
(2) efficient data movement through memory hierarchy,
(3) overlapped partitioning of data, 
(4) control over placement of data and computation,
(5) support for accelerators, and 
(6) overlapping communication and computation. 
Finally, various DSLs, such as DISTAL~\cite{DISTAL} and SpDISTAL~\cite{SpDistal}, use the Legion runtime system to implement sparse and dense tensor algebras, which allow user specification of communication patterns and of the data layout at per-node and across-nodes level, by means of scheduling languages.

\paragraph{Array Dependence Analyses}
A rich body of work has used layouts of some sort or another in the quest of optimizing affine and non-affine programs. 
%
For non-affine programs, such as molecular-dynamics simulations, inspector-executor techniques have been devised to reorder the data and iteration space at runtime~\cite{it-data-reorder-1,it-data-reorder-2}, in a way that optimizes temporal and spatial locality. For example, the inspector code computes a {\em permutation} of the data/iterations that is used by the statically-generated~executor~code.

Work on automatic parallelization of non-affine loops~\cite{pred-suif,set-congr} tests at runtime sufficient conditions for statically irreducible queries that model loop independence. These can be represented as predicated extensions of polyhedral systems~\cite {pluto,SUIF} or as languages~\cite{USR-lmad,CIVan} that build on linear-memory access descriptors (LMADs). 

LMADs~\cite{LMAD2,LMAD} generalize Python-like slicing by allowing a global-memory offset together with a list that pairs up the length of each logical dimension with its total stride---i.e., the number of memory elements that are jumped to advance to the next element in that dimension, 
similar to Graphene. 

LMADs have also been used in Futhark~\cite{futhark-mem-sc} to support various optimizations that are not expressible in a pure IR, and more relevant, to allow change-of-layout transformations to be applied on arrays at $O(1)$ cost, i.e., without manifestation in memory.  
While Figure~$3$ of~\cite{futhark-mem-sc} hints that any (straight-line) reordering sequence can be modeled by a chain of LMADs, subsequent work~\cite{futhark-mem-ifl}, presenting the memory lowering, clarifies that Futhark supports at $O(1)$ cost only reorderings that are expressible by {\em one} LMAD.

In comparison, LEGO supports reordering chains that may require several LMADs, such as \texttt{B} ($O_2$) in Figure~\ref{fig:lego-complex-eg}, and non-linear (user-defined) patterns, e.g.,  \texttt{C} ($O_1$).

\section{Conclusion} \label{sec:concl}
This paper has described LEGO, a layout algebra 
to support tiled, hierarchical high-performance code generation.  The key advance in LEGO is that it eliminates the need to specify strides in hierarchical layouts, thus simplifying layout specification.  It is also a standalone Python code that can provide a \amirbijective{bijective} mapping of computation and data layout to/from program index space, thus eliminating the need for programmers to derive complex indices manually.  It facilitates exploration of layouts in combination with other optimizations.  
We have demonstrated LEGO's integration with CUDA templates, the Triton and MLIR compilers,  and its role in generating high-performance implementations.

\amirbijective{LEGO’s support extends beyond the strided, rectangular layouts of the CuTe/Graphene shape algebra, enabling arbitrary permutations of elements, such as the anti-diagonal example and the brick data layout presented in this paper. Layout composition can be used to express data movement to and from GPU shared memory, while also supporting optimizations such as thread coarsening and the reduction of shared memory bank conflicts.}
As future work, we plan to further explore the full range of layouts and integration with other systems.

\section*{Acknowledgment}

We thank John Regehr and David Yue for their feedback on the manuscript. We are especially grateful to the anonymous reviewers for their careful reading, detailed comments, and insightful suggestions, which significantly improved the clarity and quality of this work. The support and resources from the Center for High-Performance Computing at the University of Utah are gratefully acknowledged. 
 Research reported in this publication was supported by the National Science Foundation through award CCF-2107556, and  Novo Nordisk Foundation through award NNF24OC0090447.

\appendix
\section{Artifact Appendix}

\subsection{Abstract}

This artifact contains the source code of the LEGO framework and the scripts used to execute and evaluate all benchmarks in the paper. LEGO provides an algebraic, compiler-agnostic framework for specifying and transforming memory layouts. Through integrations with Triton, CUDA, and MLIR, we compare LEGO-generated kernels with existing implementations and demonstrate that careful data layout reorganization can achieve state-of-the-art performance or significantly improve performance.

\subsection{Artifact check-list (meta-information)}


{\small
\begin{itemize}
  \item {\bf Data set:} provided as an artifact
  \item {\bf Hardware: } NVIDIA A100 80GB GPU and AMD EPYC 7513 CPU
  \item {\bf Output:}  Benchmark figures and throughput table
  \item {\bf How much disk space required (approximately)?:} 40GB
  \item {\bf How much time is needed to prepare workflow (approximately)?: } 2 hours
  \item {\bf How much time is needed to complete experiments (approximately)?: } 1.5 hours
  \item {\bf Publicly available?: } Yes
  \item {\bf Code licenses (if publicly available)?: } MIT
  \item {\bf Archived (provide DOI)?: } 10.5281/zenodo.17633994
\end{itemize}

\subsection{Description}

\subsubsection{How delivered}
10.5281/zenodo.17633994

\subsubsection{Hardware dependencies}

NVIDIA A100 80GB GPU

\subsubsection{Software dependencies}

LLVM (commit \texttt{556ec4a}), Triton 3.2.0, PyTorch 2.5.1, CUDA 12.4, Python 3.12.4, Ninja 1.12.1, CMake 3.26.5, GCC 11.2.0


\subsection{Installation}

\begin{enumerate}
  \item \textbf{Clone the LLVM}
{\scriptsize
\begin{verbatim}
$ git clone https://github.com/llvm/llvm-project.git
$ cd llvm-project && git checkout 556ec4a
\end{verbatim}
}

\item \textbf{Build and install LLVM/MLIR at the required commit}
{\scriptsize
\begin{verbatim}
$ mkdir build && cd build
$ cmake -G Ninja ../llvm -DCMAKE_BUILD_TYPE=Release \
  -DLLVM_ENABLE_PROJECTS="mlir" \
  -DLLVM_TARGETS_TO_BUILD="X86;NVPTX" \
  -DMLIR_ENABLE_CUDA_RUNNER=ON \
  -DMLIR_ENABLE_BINDINGS_PYTHON=ON \
  -DPython3_EXECUTABLE="$(which python)" \
  -DLLVM_BUILD_EXAMPLES=OFF && ninja
\end{verbatim}
}

\item \textbf{Set the path to LLVM/MLIR build path}
{\scriptsize
\begin{verbatim}
$ export MLIR_BUILD_FOLDER="$(pwd)"
\end{verbatim}
}

\end{enumerate}

\subsection{Experiment workflow}

From the root of the artifact repository, the experiments can be reproduced with the following steps:

\begin{enumerate}
  \item \textbf{Create the virtual environment and install Python packages:}

{\scriptsize
\begin{verbatim}
$ bash ./setup.sh && source venv/bin/activate
\end{verbatim}
}

  \item \textbf{Generate all kernel source code:}

{\scriptsize
\begin{verbatim}
$ bash ./gen_all_kernel.sh
\end{verbatim}
}

  \item \textbf{Run all benchmarks and produce figures and tables:}

{\scriptsize
\begin{verbatim}
$ bash ./run_all_kernels.sh
\end{verbatim}
}

This will execute all benchmarks and generate the figures and tables reported in the paper.
\end{enumerate}

\subsection{Evaluation and expected result}

The generated figures for the evaluation section (\ref{fig:eval_triton},~\ref{fig:cuda_perf},~\ref{fig:cuda-roofline}) and Table~\ref{tab:MLIR_results} are located in the \texttt{./figures} folder in the root of the artifact directory.







\balance
\bibliographystyle{IEEEtran}
\bibliography{refs}


\end{document}

%% file: prelude.tex
\usepackage{amsmath}
\usepackage{mathtools}
\usepackage{amsfonts}
\usepackage{amsthm}
\usepackage[inline]{enumitem}
\usepackage{array}
\usepackage{semantic}
\usepackage{amsthm}
\usepackage{graphicx}
\usepackage{subcaption}
\usepackage{hyperref}
\usepackage[capitalise]{cleveref}
\usepackage{listings}
\usepackage{tikz}
\usepackage{xcolor}
\usepackage{accents}

\usepackage{algorithm}
\usepackage{algorithmicx}
\usepackage[noend]{algpseudocode}

\usepackage[old]{old-arrows}

\usetikzlibrary{shapes.geometric, arrows}
\theoremstyle{definition}

\newcommand{\hsp}{\hspace{1em}}

\newcommand{\kw}[1]{\mathbf{#1}}

\newcommand{\forix}{
  \mathrm{f\kern-0.15emo\kern-0.06emr}
}

\newcommand{\scalar}{\kern-0.2em\bullet\kern-0.3em}

\newcommand{\threeqm}{\mathop{?\kern0.1em\!\raisebox{0.3em}{?}\!\kern0.1em?}}

\usepackage{soul}